\documentclass[prd,preprint,tightenlines,floatfix,showpacs,preprintnumbers,nofootinbib,eqsecnum]{revtex4-1}

\pdfoutput=1

\usepackage[dvips,final]{graphicx}
\usepackage{amssymb}
\usepackage{amsmath}
\usepackage{amsfonts}
\usepackage{epsfig}
\usepackage{bm}
\usepackage{comment}
\usepackage{float}
\usepackage[utf8]{inputenc}
\usepackage{caption}
\usepackage{subcaption}
\usepackage{hhline}

\usepackage[dvipsnames,table,xcdraw]{xcolor}
\usepackage{multirow}
\usepackage{bbold}
\usepackage{hyperref}
\hypersetup{colorlinks=true, linkcolor=teal, urlcolor=blue, citecolor=red}
\usepackage{enumerate}
\usepackage{mathtools}
\usepackage{physics} %Bunch of commands for equations
\usepackage{makecell} 

\usepackage{cancel}

\usepackage{feynmp-auto} %Feynman Diagrams

\usepackage[capitalise]{cleveref}

\crefname{section}{Sec.}{Secs.}
\crefname{table}{Tab.}{Tabs.}
\crefname{figure}{Fig.}{Figs.}
\crefname{equation}{Eq.}{Eqs.}
\crefname{appendix}{Appendix\ }{Appendix\ }

\newcommand{\U}[1]{\mathrm{U}(1)_{\mathrm{#1}}}			% Use this for U(1) groups
\definecolor{bostonuniversityred}{rgb}{0.8, 0.0, 0.0}

\newcommand\varpm{\mathbin{\vcenter{\hbox{%
  \oalign{\hfil$\scriptstyle\hspace{-0.1ex}+\hspace{-0.1ex}$\hfil\cr
          \noalign{\kern-.5ex}
          $\scriptscriptstyle({-})$\cr}%
}}}}

\DeclareSymbolFont{myletters}{OML}{ztmcm}{m}{it}
\DeclareMathSymbol{\uplambda}{\mathord}{myletters}{"15}

\begin{document}

\title{Phenomenology of a flavoured multiscalar BGL-like model with 
three generations of massive neutrinos}

\author{P.M. Ferreira$^{1,2}$}
\email{pedro.ferreira@fc.ul.pt}

\author{Felipe F. Freitas$^{3}$}
\email{felipefreitas@ua.pt}

\author{João Gonçalves$^{3}$}
\email{jpedropino@ua.pt}

\author{Antonio P. Morais$^{3}$}
\email{aapmorais@ua.pt}

\author{Roman Pasechnik$^{4}$}
\email{Roman.Pasechnik@thep.lu.se}

\author{Vasileios Vatellis$^{3}$}
\email{vasileios.vatellis@gmail.com}

\affiliation{
\\
{$^1$\sl
Instituto~Superior~de~Engenharia~de~Lisboa~---~ISEL, 1959-007~Lisboa, Portugal
}\\
{$^2$\sl
Centro~de~F\'{\i}sica~Te\'orica~e~Computacional, Faculdade~de~Ci\^encias,
Universidade~de~Lisboa, Campo Grande, 1749-016~Lisboa, Portugal
}\\
{$^3$\sl 
Departamento de F\'isica, Universidade de Aveiro and CIDMA, Campus de Santiago, 
3810-183 Aveiro, Portugal
}\\
{$^4$\sl
Department of Astronomy and Theoretical Physics, Lund
University, SE-223 62 Lund, Sweden
}
}

\begin{abstract}
In this paper, we present several possible anomaly-free implementations of the Branco-Grimus-Lavoura (BGL) model with two Higgs doublets and one singlet scalar. The model also includes three generations of massive neutrinos that get their mass  via a type-I seesaw mechanism. A particular anomaly-free realization, which we dub $\nu$BGL-1 scenario, is subjected to an extensive phenomenological analysis, from the perspective of flavour physics and collider phenomenology.
\end{abstract}

\maketitle

%%%%%%%%%%%%%%%%%%%%%%%%%%%%%%%%%%%%%%%
\section{Introduction}\label{s:intro}
%%%%%%%%%%%%%%%%%%%%%%%%%%%%%%%%%%%%%%%

With the discovery of the Higgs boson at the LHC in 2012 \cite{ATLAS:2012yve,CMS:2012qbp}, the entire particle spectrum of the Standard Model (SM) has been observed and experimentally validated. However, despite its remarkable success, there is experimental evidence for the existence of Dark Matter (DM) and neutrino masses, while the observed hierarchical structure that characterizes the fermion sector finds no explanation in the SM. With a plethora of New Physics (NP) extensions that aim at offering solutions to the aforementioned phenomena, the vast majority of them come in the form of extended scalar sectors. One of the most popular and thoroughly explored multi-scalar scenarios is the Two-Higgs doublet model (THDM), as first proposed by T.D. Lee \cite{Lee:1973iz}. The introduction of a new complex scalar doublet, in addition to the one that is already present in the SM, results in an extended particle spectrum with a new electrically charged scalar and two new neutral ones, one of which is even while the other is odd under discrete CP (charge times parity) transformations.

This type of models provide a rich breeding ground for phenomenological studies, with the possibility of having a suitable DM candidate in its spectrum \cite{CarcamoHernandez:2021iat,Arcadi:2021yyr,Bell:2016ekl,Dorsch:2017nza,Camargo:2019ukv,Ivanov:2017dad}, spontaneous CP-violation \cite{Branco:2011iw,Grzadkowski:2013rza,Boto:2020wyf,Grzadkowski:2016szj,Sokolowska:2008bt} and generation of flavour-changing neutral currents (FCNCs) mediated by the new neutral scalars \cite{Crivellin:2013wna,Kim:2015zla}. However, the presence of FCNCs can be troublesome since they are very constrained experimentally, and in the
models we are discussing they can manifest themselves already at tree-level, whereas equivalent processes in the SM are
of pure radiative origin, thus highly suppressed. As such, one either requires FCNCs couplings to be fine-tuned or rather there should exist an underlying symmetry that naturally mitigates the effect of these interactions. A concrete mechanism for symmetry-suppressed FCNCs has been proposed by Branco, Grimus and Lavoura \cite{Branco:1996bq} where, due to the presence of an abelian flavour symmetry such as $\mathbb{Z}_2$ or $\mathrm{U(1)}$, FCNC couplings are kept under control by making such interactions proportional to off-diagonal elements of the Cabibbo–Kobayashi–Maskawa (CKM) matrix. Such a class of models is commonly dubbed as BGL models.

A natural extension of the THDM-BGL framework can emerge in the form of a gauge $\mathrm{U(1)}$ symmetry, which necessarily requires a further enlargement of the scalar sector with a new EW singlet, thus falling in the family of Next-to-minimal THDM (or NTHDM) for short. Recently, a complete list of all possible anomaly-free implementations of the gauged NTHDM with two families of right-handed neutrinos and a type-I seesaw mechanism was introduced in Ref.~\cite{Ordell:2020yoq}, while a generalization to three neutrino generations was discussed in Ref.~\cite{Astrid:2019}. For previous phenomenological considerations in a set of THDMs featuring the absence of tree-level FCNCs and accommodating a seesaw mechanism for neutrino mass generation, see Ref.~\cite{Campos:2017dgc}. 
%Along these lines, there also exists previous literature which tackles this identical problem \cite{Campos:2017dgc}.
Among all allowed textures in the Yukawa sector one of them replicates the BGL structure. The model explored in this article is inspired by such a scenario but for the case of a global $\mathrm{U(1)^\prime}$ flavour symmetry. The latter is explicitly broken via soft terms in the scalar potential as well as by the vacuum expectation value (VEV) in the real component of the complex singlet scalar field. As a result, in addition to a SM-like Higgs boson candidate, five new physical scalars emerge, including an electrically charged one, $H^\pm$, two CP-even neutral ones, $H_2$ and $H_3$, and two pseudoscalars, $A_2$ and $A_3$. Note that in the context of the gauge $\mathrm{U(1)^\prime}$ flavour symmetry \cite{Ordell:2020yoq, Astrid:2019} one of those pseudoscalars, as a pure Goldstone particle, becomes a longitudinal polarization mode of a $Z^\prime$ vector boson. In turn, tight constraints imposed by direct searches at the LHC \cite{ATLAS:2020lks,ATLAS:2018nda,ATLAS:2021shl,CMS:2019gwf,CMS:2021ctt,CMS:2018hnz} on new gauge bosons imply a rather large $\mathrm{U(1)}^\prime$ breaking scale, of at least a few TeV, naturally inducing masses of the same order to the new scalars, unless a certain degree of fine-tuning is in place. Our current goal is to study the impact of the extended scalar sectors not too far from the EW scale. For this purpose, a global version of the BGL scenario instead of the gauged one proposed in Ref.~\cite{Ordell:2020yoq, Astrid:2019} is considered in this work. 

We test the viability of the scalar sector against phenomenological observables, in particular, by making sure that FCNCs obey current experimental constraints and that there exists one CP-even neutral scalar that replicates the properties of the SM Higgs boson, defined as $H_1$ in this article. For a rigorous analysis we employ widely used open-source software, namely, \texttt{SARAH} \cite{Staub:2013tta}, for model implementation, and \texttt{SPheno} \cite{Porod:2011nf}, to numerically calculate masses and interaction vertices, which are then used to interface with \texttt{HiggsBounds} and \texttt{HiggsSignals} \cite{Bechtle:2015pma,Bechtle:2013xfa}, where the viability of the Higgs sector is verified, and \texttt{flavio} \cite{Straub:2018kue}, where flavour observables are calculated.

The paper is organised as follows. In Sec.~\ref{s:lag} we introduce the model, with the focus on the Yukawa and scalar sectors of the theory that obey the BGL structure. In Sec.~\ref{s:procedure} we detail the method by which one can obtain anomaly-free models, as first discussed in Ref.~\cite{Astrid:2019}, highlighting the possible textures in the lepton sector. In Sec.~\ref{s:AnomaFree} we show all possible anomaly-free implementations of the considered class of THDSM scenarios. In Sec.~\ref{sec:chosen_scenario} we discuss the particular version of the model considered in this article, with a discussion on the neutrino masses and mixings. In Sec.~\ref{sec:Pheno_section} we describe the methodology behind the phenomenological studies and present the numerical results that follow. In Sec.~\ref{sec:LHC} we confront our results with direct scalar searches at the LHC. In Sec.~\ref{sec:conclusions} we conclude with a summary of our findings.

%%%%%%%%%%%%%%%%%%%%%%%%%%%%%%%%%%%%
\section{A generic NTHDM with a BGL structure}\label{s:lag}
%%%%%%%%%%%%%%%%%%%%%%%%%%%%%%%%%%%%

In this work, we extend the SM with a flavour non-universal $\U{}^\prime$ global symmetry alongside three generations of right-handed neutrinos $\nu_R^{1,2,3}$, a scalar singlet $S$ and a second Higgs doublet\footnote{The BGL model versions discussed in this article were first introduced in the Appendix of \cite{Astrid:2020}.} $\Phi_2$. In addition, we demand that the singlet and at least one of the doublets transform non-trivially under $\mathrm{U(1)}^\prime$ such that at least two VEVs contribute to the breaking of the flavor symmetry.

The Yukawa interactions that define the model to be discussed read as
\begin{align}
\label{YUK}
\begin{split}
-\mathcal{L}_\mathrm{Yukawa}&=\overline{q_L^0}\Gamma_a\Phi^a d_R^0+\overline{q_L^0}\Delta_a\tilde{\Phi}^a u_R^0
+\overline{\ell_L^0}\Pi_a\Phi^a e_R^0+\overline{\ell_L^0}\Sigma_a\tilde{\Phi}^a \nu_R^0\\
&+\frac{1}{2}\overline{\nu_R^{c\,0}}\left(\mathrm{A}+\mathrm{B}S+\mathrm{C}S^\ast\right)\nu_R^0+\mathrm{h.c.}\,,
\end{split}
\end{align}
where $\Gamma$, $\Delta$, $\Pi$ and $\Sigma$ are the $3\times3$ Dirac-like down quark, up quark, charged lepton and neutrino Yukawa coupling matrices, respectively. The index $a$ runs over the two scalar doublets (with implicit summation) and, as usual, $\tilde{\Phi}\equiv i \sigma_2{\Phi}^*$. While B and C are the Majorana-like Yukawa matrices involving right-handed neutrinos and the scalar singlet, A is a Majorana mass term written in the flavour basis, defined by the fields expressed with the $0$ superscript. To comply with a BGL structure the transformation laws of the 
different fields under the $\U{}^\prime$ global symmetry are chosen so that the quark Yukawa textures are given as
\begin{align}
\label{BGLtextures}
\begin{split}
\Gamma_1&:
\begin{pmatrix}
\times& \times & \times
\\
\times &\times & \times
\\
 \makebox[\widthof{$\times$}][c]{0} & \makebox[\widthof{$\times$}][c]{0}&  \makebox[\widthof{$\times$}][c]{0}
\end{pmatrix},
\hspace{2mm}
\Gamma_2:
\begin{pmatrix}
 \makebox[\widthof{$\times$}][c]{0}&  \makebox[\widthof{$\times$}][c]{0}& \makebox[\widthof{$\times$}][c]{0}
\\
 \makebox[\widthof{$\times$}][c]{0}&    \makebox[\widthof{$\times$}][c]{0}& \makebox[\widthof{$\times$}][c]{0}
\\
\times & \times & \times
\end{pmatrix},
\hspace{2mm}
\\
\Delta_1&:
\begin{pmatrix}
\times& \times &\makebox[\widthof{$\times$}][c]{0}
\\
\times&\times & \makebox[\widthof{$\times$}][c]{0}
\\
\makebox[\widthof{$\times$}][c]{0} &\makebox[\widthof{$\times$}][c]{0} &\makebox[\widthof{$\times$}][c]{0}
\end{pmatrix},
\hspace{1,5mm}
\Delta_2:
\begin{pmatrix}
\makebox[\widthof{$\times$}][c]{0} &\makebox[\widthof{$\times$}][c]{0} & \makebox[\widthof{$\times$}][c]{0} 
\\
\makebox[\widthof{$\times$}][c]{0}&\makebox[\widthof{$\times$}][c]{0}  & \makebox[\widthof{$\times$}][c]{0}
\\
\makebox[\widthof{$\times$}][c]{0} & \makebox[\widthof{$\times$}][c]{0}& \times
\end{pmatrix}.
\end{split}
\end{align}
Note that this choice of textures implies that tree-level FCNCs will appear only in the down quark sector, since the up-quark Yukawa matrices can be diagonalized simultaneously. Indeed, this represents the more conservative scenario, since down-type FCNCs are more tightly constrained by experiment when compared to those in the up-sector. Of course, one can also extend the BGL mechanism to the lepton sector, as first proposed in Ref.~\cite{Botella:2011ne}, which is beyond the scope of the present work. 

In Eq.~\eqref{YUK}, the scalar fields can be expanded as
\begin{align}
\begin{split}
\label{eq:Field_ext}
\Phi_a &\equiv \frac{1}{\sqrt{2}}
\begin{pmatrix}
\sqrt{2}\phi^+_a\\
v_a\,e^{i\varphi_a}+R_a+iI_a
\end{pmatrix}\,,\\[0.8mm]
S &\equiv\frac{1}{\sqrt{2}} \left(v_S\,e^{i\varphi_S}+\rho+i\eta\right)\,,
\end{split}
\end{align} 
where $v_{1,2}$ and $v_S$ denote the doublet and singlet VEVs, respectively. As the scalar fields have, in general, different $\mathrm{U(1)_Y \times U(1)^\prime}$ charge assignment, we can choose two of the VEVs real, for example $v_{1}$ and $v_2$, i.e.~$\varphi_{1,2}=0$. In multi-Higgs SM extensions there is a possibility to trigger spontaneous CP breaking via complex VEVs. For the purposes of this work, however, we are only focused on a CP conserving scalar phase. CP 
violation will emerge via the CKM matrix as in the SM, due to the fact that the entries of the Yukawa matrices
of equation~\eqref{BGLtextures} are complex.

Once the two doublets acquire VEVs, both the quarks and the charged leptons obtain tree-level masses which are given by
\begin{equation}\label{eq:tree_level_masses}
\begin{aligned}
&M_u^0\equiv\frac{1}{\sqrt{2}}\left(v_1 \Delta_1+v_2\Delta_2\right), \\
&M_d^0=\frac{1}{\sqrt{2}}\left(v_1 \Gamma_1+v_2\Gamma_2\right), \\
&M_e^0=\frac{1}{\sqrt{2}}\left(v_1 \Pi_1+v_2\Pi_2\right).
\end{aligned}
\end{equation}
One can rotate the spectrum to the mass basis via bi-unitary transformations of the form 
\begin{equation}\label{eq:biunitary_fermions}
D_f = U_{f\mathrm{L}}^\dagger M_f^0 \, U_{f\mathrm{R}},
\end{equation}
with $D_f$ a diagonal mass form with ordered fermion masses and where $U_\mathrm{f\,L,R}$ are unitary matrices with the subscript $f$ indicating the fermion flavour, \textit{i.e.}~$f = e,u,d$. Following the standard notation, one can define the following matrices
\begin{align}
N_u^0=\frac{1}{\sqrt{2}}\big(v_2 {\Delta_{1}}-v_1{\Delta_{2}}\big)\, \quad\quad
N_d^0=\frac{1}{\sqrt{2}}\big(v_2 {\Gamma_{1}}-v_1{\Gamma_{2}}\big)\,,
\end{align}
whose off-diagonal elements are responsible for inducing tree-level FCNC interactions. The key feature of the BGL model is that such matrices can be re-expressed solely in terms of quark masses, CKM mixing elements $V_{ij} \equiv \sum_k (U_{u\mathrm{L}})_{ik} (U_{d\mathrm{L}}^\dagger)_{kj}$, with $i,j,k = 1,2,3$, and the ratio $t_\beta \equiv \tan\beta \equiv v_1/v_2$ as follows
\begin{align}
\begin{split}
\left(N_u\right)_{ij}&=\left(t_\beta\delta_{ij}-\left(t_\beta+t_\beta^{-1}\right)\delta_{ij}\delta_{j3}\right)m_{u_j},
\\
\left(N_d\right)_{ij}&=\left(t_\beta\delta_{ij}-\left(t_\beta+t_\beta^{-1}\right)V_{3i}^*V_{3j}\right)m_{d_j}\,,
\end{split}
\end{align}
such that the down-specific FCNCs are controlled by the smallness of the off-diagonal CKM mixing elements $V_{3i}$.

For the charged lepton sector, the generic coupling combinations can be found in complete analogy with the quarks, i.e.
\begin{align}
M_e^0=\frac{1}{\sqrt{2}}\left(v_1\Pi_1+v_2\Pi_2\right)\,,\quad\quad
N_e^0=\frac{1}{\sqrt{2}}\left(v_2\Pi_1-v_1\Pi_2\right)\,.
\end{align}
Last but not least, the neutrino sector is equipped with a seesaw mechanism of type-I, as soon as the singlet $S$ develops a VEV. In particular, defining 
\begin{align}
n_L^0\equiv
\begin{pmatrix}
\nu_L^0\\
\nu_R^c
\end{pmatrix},
\end{align}
one can recast the neutrino mass Lagrangian as
\begin{align}
\begin{split}
-\mathcal{L}_{\nu}^\mathrm{mass}=&\frac{1}{2}\overline{n_L^0}\mathcal{M} n_{L}^{0,c}+\text{h.c.}\,,
\end{split}
\end{align}
where
\begin{align}
\mathcal{M}\equiv
\begin{pmatrix}
\mathbb{0}&m_D\\
m_D^T&M_R
\end{pmatrix},
\end{align}
and with the Dirac and Majorana neutrino masses given by
\begin{align}
\begin{split}
m_D&\equiv\frac{1}{\sqrt{2}}\left(v_1\Sigma_1+v_2\Sigma_2\right),\\
M_R&\equiv\mathrm{A}+\frac{v_S}{\sqrt{2}}\left(\mathrm{B}+\mathrm{C}\right)\,,
\end{split}
\end{align}
respectively. Here, we have also considered that the singlet VEV is real. At low energies, in the limit of $M_R \gg m_D/m_D^T$, the effective Lagrangian for the light active neutrinos takes the form
\begin{align}
\begin{split}
-\mathcal{L}_\nu^\mathrm{eff}\equiv\frac{1}{2}\overline{\nu_L^0}m_\nu \nu_L^{0,c}+\text{h.c.},
\end{split}
\end{align}
with
\begin{align}
m_\nu\equiv-m_D M_R^{-1} m_D^T\,.
\end{align}
The numerical analysis performed in this study relies on an inversion procedure analogous to that developed in \cite{Das:2021oik}. In particular, for the quark sector, one uses their physical masses as well as the CKM mixing elements as input parameters. However, due to a rich neutrino content, one cannot analytically invert the entire lepton sector. Instead, while electron, muon and tau masses are given as input parameters, one performs a numerical fit to neutrino masses alongside an extended version of the Pontecorvo-Maki-Nakagawa-Sakata (PMNS) matrix, as it is discussed in \cref{sec:chosen_scenario}.

The most generic form of the scalar potential can be written as
\begin{align}\label{eq:Scalar_pot}
    V = V_0 + V_1
\end{align} 
with
\begin{align}
\label{Scal_Pote_1}
\begin{split}
V_0 =\quad  & \mu_{i}^2|\Phi^{i}|^2 + \lambda_{i}|\Phi^{i}|^4 + \lambda_{3}|\Phi_{1}|^2|\Phi_{2}|^2 + \lambda_{4}|\Phi_{1}^{\dagger}\Phi_{2}|^2  + {\mu_S}^{2}|S|^2 +  \lambda_{1}^\prime|S|^4 \\
& + \lambda_{2}^\prime|\Phi_{1}|^2|S|^2 + \lambda_{3}^\prime|\Phi_{2}|^2|S|^2 \qquad \textrm{and}\\
V_1 =\quad & \mu_{3}^{2} \Phi_{2}^{\dagger}\Phi_1  + \frac{1}{2}\mu_{b}^{2} S^{2} + a_{1} \Phi_{1}^{\dagger}\Phi_2 S + a_{2} \Phi_{1}^{\dagger}\Phi_2 S^{\dagger} +  a_{3} \Phi_{1}^{\dagger}\Phi_2 S^{2} + a_{4} \Phi_{1}^{\dagger}\Phi_2 {S^{\dagger}}^2 + \mathrm{h.c.}\, .
\end{split}
\end{align}
While in $V_0$ we adopt the usual notation where $\mu_{i,S}$ represent quadratic mass parameters and $\lambda_i$, $\lambda_i^\prime$ are the quartic couplings, in $V_1$ $a_{1,2}$ denote cubic couplings between the Higgs doublets and the singlet while $a_{3,4}$ are the quartic couplings between the two Higgs doublets and the scalar singlet field. Given that the singlet $S$ carries a non-trivial $\U{}^\prime$ charge $X_S$, then, out of the four $a_{1,2,3,4}$ terms, only one is allowed in the limit of an exact $\U{}^\prime$. However, both $a_1$ and $a_2$, as well as $\mu_b^2$, can be introduced to softly break the flavour symmetry and are allowed to coexist with either $a_3$ or $a_4$~\cite{Astrid:2019,Ordell:2020yoq}.

%%%%%%%%%%%%%%%%%%%%%%%%%%%%%%%%%%%%
\section{Anomaly-free BGL implementation}\label{s:procedure} 
%%%%%%%%%%%%%%%%%%%%%%%%%%%%%%%%%%%%

The transformation properties of the quark and scalar fields under the $\U{}^\prime$ flavor symmetry are given, as usual,  by
\begin{equation}\label{eq:fields_trans}
\Psi_i \rightarrow e^{i\theta X_{\Psi_i}}\Psi_i \,,
\end{equation}
where $\Psi_i \subset (q_L^0, u_R^0, d_R^0, \Phi_1, \Phi_2, S)$, and where $X_{\Psi_i}$ are the corresponding 
$\mathrm{U(1)}^\prime$ charges with $\theta$ a global phase parameter. It then follows that the Yukawa matrices transform as
\begin{align}
\label{eq:SymmetryTextures}
\begin{split}
(\Gamma_a)_{ij} &= e^{i\theta(X_{q_i}-X_{d_j}- X_{\Phi_a})}(\Gamma_a)_{ij}\,,
\\
(\Delta_a)_{ij} &= e^{i\theta(X_{q_i}-X_{u_j}+ X_{\Phi_a})}(\Delta_a)_{ij}\,, 
\end{split}
\end{align}
such that textures on $(\Gamma_a)_{ij}$ result 
from a set of linear constraints that read as
\begin{align}
\label{eq:ResTextures}
\begin{split}
(\Gamma_a)_{ij}&=\mathrm{any\;\;if}\;\;X_{q_i}-X_{d_j}= X_{\Phi_a}\,, 
\\
(\Gamma_a)_{ij}&=0\;\;\;\;\;\;\mathrm{if}\;\;X_{q_i}-X_{d_j}\neq X_{\Phi_a}\,.
\end{split}
\end{align}
The same can be done in the up-quark sector by trading $(\Gamma_a)_{ij} \to (\Delta_a)_{ij}$, $X_{d_j} \to X_{u_j}$ 
and $X_{\Phi_a} \to -X_{\Phi_a}$. Among many possible solutions found in Refs.~\cite{Astrid:2019} and \cite{Ordell:2020yoq}, 
the well-known quark BGL textures (see Eq.~\eqref{BGLtextures}), that are used in this work, 
satisfy the following 36 constraints
\begin{align}
\begin{split}
&X_{q_{1,2}}-X_{d_{1,2,3}}=X_{\Phi_1}, \;\;X_{q_{3}}-X_{d_{1,2,3}}\neq X_{\Phi_1} \,,
\\
&X_{q_{3}}-X_{d_{1,2,3}}=X_{\Phi_2},\;\;X_{q_{1,2}}-X_{d_{1,2,3}}\neq X_{\Phi_2} \,,
\\
&X_{q_{1,2}}-X_{u_{1,2}}=-X_{\Phi_1}, \;\;X_{q_{3}}-X_{u_{1,2,3}}\neq -X_{\Phi_1} \,,
\\
&X_{q_{1,2}}-X_{u_{3}}\neq -X_{\Phi_1},\;\;X_{q_{3}}-X_{u_{3}}=-X_{\Phi_2} \,,
\\
&X_{q_{1,2}}-X_{u_{1,2,3}}\neq -X_{\Phi_2},\;\;X_{q_{3}}-X_{u_{1,2}}\neq -X_{\Phi_2} \,.
\end{split}
\end{align}
Focusing now on the lepton and neutrino sectors, one must consider all possible textures of the Dirac-like $\Pi_{a}$ and $\Sigma_{a}$ 
and Majorana-like $B$ and $C$ Yukawa matrices, as well as on the $A$ mass matrix. To do so one considers that
\begin{itemize}
\item[(i)] there are no massless charged leptons yielding $\text{det}\hspace{0.5mm}M_e\neq 0$;
\item[(ii)] there are three generations of massive neutrinos such that\footnote{Note that there exists no anomaly-free implementation of the BGL textures with two generations of massive neutrinos, as shown in Ref.~\cite{Ordell:2020yoq}.} $\text{det}\hspace{0.5mm}M_\nu\neq 0$;
\item[(iii)] there is a non-zero complex phase in the PMNS matrix which implies that\\ $\text{det}\hspace{0.5mm}[M_eM_e^\dagger] \neq 0$ and $\text{det}\hspace{0.5mm}[M_\nu M_\nu^\dagger]\neq 0$\,.
\end{itemize}
Notice that the second condition is the only one differing from those first introduced in Ref.~\cite{Astrid:2019} and \cite{Ordell:2020yoq}. For a type-I seesaw mechanism such a condition also translates into $M_\mathrm{R}$ and $M_\mathrm{D}$ being $3\times3$ matrices with non-zero determinant. Note also that any two models that can be mapped from one another via a permutation are equivalent as it would simply correspond to a relabelling of flavor indices. Defining the new primed Yukawa and mass matrices as
\begin{align}
\label{permutations}
\begin{split}
&\Gamma^\prime_{1,2}=\mathcal{P}_i^\mathrm{T}\Gamma_{1,2} \mathcal{P}_j \,, \;\;\Delta^\prime_{1,2}=\mathcal{P}_i^\mathrm{T}\Delta_{1,2} \mathcal{P}_k \,,
\\
&\Pi^\prime_{1,2}=\mathcal{P}_l^\mathrm{T}\Pi_{1,2} \mathcal{P}_m \,, \;\;\Sigma^\prime_{1,2}=\mathcal{P}_l^\mathrm{T}\Sigma_{1,2} {\mathcal{P}}_n \,,
\\
&A^\prime={\mathcal{P}}_n^\mathrm{T} A\hspace{0.4mm}{\mathcal{P}}_n \,, \;\;
B^\prime={\mathcal{P}}_n^\mathrm{T} B\hspace{0.4mm} {\mathcal{P}}_n \,,\;\;
C^\prime={\mathcal{P}}_n^\mathrm{T} C\hspace{0.4mm} {\mathcal{P}}_n \,,
\end{split}
\end{align}
with $\mathcal{P}$ being a three-dimensional representation of the permutation group $S_3$, with all indices 
running from one to six, one excludes all such additional textures that can be obtained by permutations 
in the flavour space.

\subsection*{Majorana neutrino sector}

In total, there are 11 minimal textures for $A$, $B$ and $C$ that fulfil the constraint of $M_\mathrm{R}$ being a $3\times3$ symmetric matrix with a non-zero determinant. These are
\begin{align}
\label{MajText}
\begin{split}
(1)&\;\;\;\;A:
\begin{pmatrix}
\times & \makebox[\widthof{$\times$}][c]{0}&  \makebox[\widthof{$\times$}][c]{0}
\\
\makebox[\widthof{$\times$}][c]{0} & \makebox[\widthof{$\times$}][c]{0} & \times
\\
\makebox[\widthof{$\times$}][c]{0} & \times & \makebox[\widthof{$\times$}][c]{0} 
\end{pmatrix},
\;\;
B:
\mathbb{0},
\;\;
C:
\mathbb{0},
\\
(2)&\;\;\;\;A:
\begin{pmatrix}
\times & \makebox[\widthof{$\times$}][c]{0}& \makebox[\widthof{$\times$}][c]{0}
\\
\makebox[\widthof{$\times$}][c]{0}& \makebox[\widthof{$\times$}][c]{0}& \makebox[\widthof{$\times$}][c]{0}
\\
\makebox[\widthof{$\times$}][c]{0}& \makebox[\widthof{$\times$}][c]{0}& \makebox[\widthof{$\times$}][c]{0}
\end{pmatrix},
\;\;
B:
\begin{pmatrix}
\makebox[\widthof{$\times$}][c]{0} & \makebox[\widthof{$\times$}][c]{0}&  \makebox[\widthof{$\times$}][c]{0}
\\
\makebox[\widthof{$\times$}][c]{0} & \makebox[\widthof{$\times$}][c]{0} & \times
\\
\makebox[\widthof{$\times$}][c]{0} & \times & \makebox[\widthof{$\times$}][c]{0} 
\end{pmatrix},
\;\;
C:
\mathbb{0},
\\[1mm]
(3)&\;\;\;\;A:
\begin{pmatrix}
\times & \makebox[\widthof{$\times$}][c]{0}& \makebox[\widthof{$\times$}][c]{0}
\\
\makebox[\widthof{$\times$}][c]{0}& \makebox[\widthof{$\times$}][c]{0}& \makebox[\widthof{$\times$}][c]{0}
\\
\makebox[\widthof{$\times$}][c]{0}& \makebox[\widthof{$\times$}][c]{0}& \makebox[\widthof{$\times$}][c]{0}
\end{pmatrix},
\;\;
B:
\begin{pmatrix}
\makebox[\widthof{$\times$}][c]{0}& \makebox[\widthof{$\times$}][c]{0}& \makebox[\widthof{$\times$}][c]{0}
\\
\makebox[\widthof{$\times$}][c]{0}& \times & \makebox[\widthof{$\times$}][c]{0}
\\
\makebox[\widthof{$\times$}][c]{0}& \makebox[\widthof{$\times$}][c]{0}& \makebox[\widthof{$\times$}][c]{0}
\end{pmatrix},\;\;C:
%\\
%&\;\;\;\;C:
\begin{pmatrix}
\makebox[\widthof{$\times$}][c]{0}& \makebox[\widthof{$\times$}][c]{0}& \makebox[\widthof{$\times$}][c]{0}
\\
\makebox[\widthof{$\times$}][c]{0}& \makebox[\widthof{$\times$}][c]{0} & \makebox[\widthof{$\times$}][c]{0}
\\
\makebox[\widthof{$\times$}][c]{0}& \makebox[\widthof{$\times$}][c]{0}& \times
\end{pmatrix},
\\[1mm]
(4)&\;\;\;\;A:
\begin{pmatrix}
\makebox[\widthof{$\times$}][c]{0}& \times& \makebox[\widthof{$\times$}][c]{0}
\\
\times& \makebox[\widthof{$\times$}][c]{0}& \makebox[\widthof{$\times$}][c]{0}
\\
\makebox[\widthof{$\times$}][c]{0}& \makebox[\widthof{$\times$}][c]{0}& \makebox[\widthof{$\times$}][c]{0}
\end{pmatrix},
\;\;
B:
\begin{pmatrix}
\makebox[\widthof{$\times$}][c]{0}& \makebox[\widthof{$\times$}][c]{0}& \times
\\
\makebox[\widthof{$\times$}][c]{0}&\makebox[\widthof{$\times$}][c]{0}& \makebox[\widthof{$\times$}][c]{0}
\\
\times& \makebox[\widthof{$\times$}][c]{0}& \makebox[\widthof{$\times$}][c]{0}
\end{pmatrix},
\;\;C:
\begin{pmatrix}
\makebox[\widthof{$\times$}][c]{0}& \makebox[\widthof{$\times$}][c]{0}& \makebox[\widthof{$\times$}][c]{0}
\\
\makebox[\widthof{$\times$}][c]{0}& \makebox[\widthof{$\times$}][c]{0} & \times
\\
\makebox[\widthof{$\times$}][c]{0}& \times& \makebox[\widthof{$\times$}][c]{0}
\end{pmatrix},
\end{split}
\end{align}
where textures (1) and (2) come in three and six versions, respectively,  taking into account all possible permutations of $A$, $B$ 
and $C$. For textures (3) and (4), on the other hand, we only need to consider those shown above as 
permutations of rows and columns solely correspond to a relabelling of flavor indices. 
In the presence of the $U(1)^\prime$ flavour symmetry one must also consider the transformation laws
\begin{equation}
\begin{aligned}
A_{ij}&=e^{i\alpha(X_{\nu_i}+X_{\nu_j})}A_{ij},
\\
B_{ij}&=e^{i\alpha(X_{\nu_i}+X_{\nu_j}+X_S)}B_{ij},
\\
C_{ij}&=e^{i\alpha(X_{\nu_i}+X_{\nu_j}-X_S)}C_{ij}.
\end{aligned}
\end{equation}
Therefore, texture (1) implies that
\begin{equation}
\label{constraintMaj}
2X_{\nu_1}=0, \;\;X_{\nu_2}+X_{\nu_3}=0,
\end{equation}
while texture (4) results in
\begin{equation}
\begin{aligned}
X_{\nu_1}+X_{\nu_2}&=0, \;\;X_{\nu_1}+X_{\nu_3}=X_S,
\\
X_{\nu_2}+X_{\nu_3}&=-X_S,
\end{aligned}
\end{equation}
and so on. In addition to this, one of the following four conditions (see Ref.~\cite{Ordell:2020yoq})
\begin{equation}
\label{ScalarPot}
\begin{aligned}
X_S=\pm\left(X_{\Phi_1}-X_{\Phi_2}\right), \;\;X_S=\pm\frac{1}{2}\left(X_{\Phi_1}-X_{\Phi_2}\right),
\end{aligned}
\end{equation}
can be extracted from the scalar potential $V_1$ depending on which of the $a_{1,2,3,4}$ terms is invariant under the flavour $\U{}^\prime$ symmetry. However, if only $A$ has a non-zero texture then $S$ is neutral under the flavour symmetry and the four conditions above are all automatically valid. In that case the cubic $a_{1,2}$ and quartic $a_{3,4}$ can simultaneously coexist.

\subsection*{Charged lepton sector}

For the charged lepton sector, the same minimal textures already discussed in Ref.~\cite{Astrid:2019} and \cite{Ordell:2020yoq} apply, \textit{i.e.}
\begin{align}
\label{texturesLepton}
\begin{split}
(1)\;\;\;\;\;\;\Pi_1:
\begin{pmatrix}
\times &   \makebox[\widthof{$\times$}][c]{0}&  \makebox[\widthof{$\times$}][c]{0}
\\
 \makebox[\widthof{$\times$}][c]{0} &   \times&   \makebox[\widthof{$\times$}][c]{0}
\\
 \makebox[\widthof{$\times$}][c]{0} &   \makebox[\widthof{$\times$}][c]{0}&    \times
\end{pmatrix},
\;\;
\Pi_2:
\begin{pmatrix}
 \makebox[\widthof{$\times$}][c]{0}&   \makebox[\widthof{$\times$}][c]{0}&  \makebox[\widthof{$\times$}][c]{0}
\\
  \makebox[\widthof{$\times$}][c]{0} &  \makebox[\widthof{$\times$}][c]{0}&   \makebox[\widthof{$\times$}][c]{0}
\\
 \makebox[\widthof{$\times$}][c]{0} &   \makebox[\widthof{$\times$}][c]{0}&   \makebox[\widthof{$\times$}][c]{0}
\end{pmatrix},
\\
(2)\;\;\;\;\;\;\Pi_1:
\begin{pmatrix}
 \times &   \makebox[\widthof{$\times$}][c]{0}&  \makebox[\widthof{$\times$}][c]{0}
\\
  \makebox[\widthof{$\times$}][c]{0} &   \times&   \makebox[\widthof{$\times$}][c]{0}
\\
 \makebox[\widthof{$\times$}][c]{0} &   \makebox[\widthof{$\times$}][c]{0}&    \makebox[\widthof{$\times$}][c]{0}
\end{pmatrix},
\;\;
\Pi_2:
\begin{pmatrix}
 \makebox[\widthof{$\times$}][c]{0}&   \makebox[\widthof{$\times$}][c]{0}&  \makebox[\widthof{$\times$}][c]{0}
\\
  \makebox[\widthof{$\times$}][c]{0} & \makebox[\widthof{$\times$}][c]{0}&   \makebox[\widthof{$\times$}][c]{0}
\\
 \makebox[\widthof{$\times$}][c]{0} &   \makebox[\widthof{$\times$}][c]{0}&   \times
\end{pmatrix},
\\
(3)\;\;\;\;\;\;\Pi_1:
\begin{pmatrix}
 \times &   \makebox[\widthof{$\times$}][c]{0}&  \makebox[\widthof{$\times$}][c]{0}
\\
  \makebox[\widthof{$\times$}][c]{0} & \makebox[\widthof{$\times$}][c]{0}&   \makebox[\widthof{$\times$}][c]{0}
\\
 \makebox[\widthof{$\times$}][c]{0} &   \makebox[\widthof{$\times$}][c]{0}&    \makebox[\widthof{$\times$}][c]{0}
\end{pmatrix},
\;\;
\Pi_2:
\begin{pmatrix}
 \makebox[\widthof{$\times$}][c]{0}&   \makebox[\widthof{$\times$}][c]{0}&  \makebox[\widthof{$\times$}][c]{0}
\\
  \makebox[\widthof{$\times$}][c]{0} &  \times&   \makebox[\widthof{$\times$}][c]{0}
\\
 \makebox[\widthof{$\times$}][c]{0} &   \makebox[\widthof{$\times$}][c]{0}&   \times
\end{pmatrix},
\\
(4)\;\;\;\;\;\;\Pi_1:
\begin{pmatrix}
 \makebox[\widthof{$\times$}][c]{0} &   \makebox[\widthof{$\times$}][c]{0}&  \makebox[\widthof{$\times$}][c]{0}
\\
  \makebox[\widthof{$\times$}][c]{0} & \makebox[\widthof{$\times$}][c]{0}&   \makebox[\widthof{$\times$}][c]{0}
\\
 \makebox[\widthof{$\times$}][c]{0} &   \makebox[\widthof{$\times$}][c]{0}&    \makebox[\widthof{$\times$}][c]{0}
\end{pmatrix},
\;\;
\Pi_2:
\begin{pmatrix}
  \times &   \makebox[\widthof{$\times$}][c]{0}&  \makebox[\widthof{$\times$}][c]{0}
\\
  \makebox[\widthof{$\times$}][c]{0} &  \times&   \makebox[\widthof{$\times$}][c]{0}
\\
 \makebox[\widthof{$\times$}][c]{0} &   \makebox[\widthof{$\times$}][c]{0}&   \times
\end{pmatrix}.
\end{split}
\end{align}

\subsection*{Dirac neutrino sector}

For the Dirac neutrino textures, permutation of rows and columns are no longer independent from those in the charged lepton and Majorana neutrino sectors. As a result, besides fulfilling the constraint of $M_\mathrm{D}$ having a non-zero determinant, we must also consider textures that are equivalent up to permutations. In total, this amounts to six possible minimal combined textures,
\begin{align*}
\begin{split}
1:
\begin{pmatrix}
\times&\makebox[\widthof{$\times$}][c]{0}&\makebox[\widthof{$\times$}][c]{0}
\\
\makebox[\widthof{$\times$}][c]{0} & \times&\makebox[\widthof{$\times$}][c]{0}
\\
\makebox[\widthof{$\times$}][c]{0} & \makebox[\widthof{$\times$}][c]{0}& \times
\end{pmatrix},
\;\;
2:
\begin{pmatrix}
\times&\makebox[\widthof{$\times$}][c]{0}&\makebox[\widthof{$\times$}][c]{0}
\\
\makebox[\widthof{$\times$}][c]{0} & \makebox[\widthof{$\times$}][c]{0}& \times
\\
\makebox[\widthof{$\times$}][c]{0} & \times& \makebox[\widthof{$\times$}][c]{0}
\end{pmatrix},
\;\;
3:
\begin{pmatrix}
\makebox[\widthof{$\times$}][c]{0} & \times&\makebox[\widthof{$\times$}][c]{0}
\\
 \times & \makebox[\widthof{$\times$}][c]{0}& \makebox[\widthof{$\times$}][c]{0}
\\
\makebox[\widthof{$\times$}][c]{0} & \makebox[\widthof{$\times$}][c]{0}& \times
\end{pmatrix},
\\
4:
\begin{pmatrix}
\makebox[\widthof{$\times$}][c]{0}&\times& \makebox[\widthof{$\times$}][c]{0}
\\
 \makebox[\widthof{$\times$}][c]{0}& \makebox[\widthof{$\times$}][c]{0}&\times 
\\
\times&\makebox[\widthof{$\times$}][c]{0} & \makebox[\widthof{$\times$}][c]{0}
\end{pmatrix},
\;\;
5:
\begin{pmatrix}
\makebox[\widthof{$\times$}][c]{0}&\makebox[\widthof{$\times$}][c]{0}& \times
\\
\times&\makebox[\widthof{$\times$}][c]{0} & \makebox[\widthof{$\times$}][c]{0}
\\
 \makebox[\widthof{$\times$}][c]{0}&\times &\makebox[\widthof{$\times$}][c]{0}
\end{pmatrix},
\;\;
6:
\begin{pmatrix}
\makebox[\widthof{$\times$}][c]{0}&\makebox[\widthof{$\times$}][c]{0}& \times
\\
\makebox[\widthof{$\times$}][c]{0} & \times&\makebox[\widthof{$\times$}][c]{0}
\\
 \times &\makebox[\widthof{$\times$}][c]{0}&\makebox[\widthof{$\times$}][c]{0}
\end{pmatrix},
\end{split}
\end{align*}
which in turn corresponds to 48 possible textures for $\Sigma_{1,2}$ -- eight for each of the textures displayed above; 111, 112, 121, 211, 122, 212, 221 and 222, where the numbers indicate the corresponding non-zero entry in $\Sigma_1$ or $\Sigma_2$. As an example, the eight possibilities for texture number 2 are given by

\begin{align*}
\begin{split}
(111)\;\;\;\;\;\;\Sigma_1:\begin{pmatrix}
\times&\makebox[\widthof{$\times$}][c]{0}&\makebox[\widthof{$\times$}][c]{0}
\\
\makebox[\widthof{$\times$}][c]{0} & \makebox[\widthof{$\times$}][c]{0}& \times
\\
\makebox[\widthof{$\times$}][c]{0} & \times& \makebox[\widthof{$\times$}][c]{0}
\end{pmatrix},
\;\;\Sigma_2:\begin{pmatrix}
\makebox[\widthof{$\times$}][c]{0}&\makebox[\widthof{$\times$}][c]{0}&\makebox[\widthof{$\times$}][c]{0}
\\
\makebox[\widthof{$\times$}][c]{0} & \makebox[\widthof{$\times$}][c]{0}& \makebox[\widthof{$\times$}][c]{0}
\\
\makebox[\widthof{$\times$}][c]{0} & \makebox[\widthof{$\times$}][c]{0}& \makebox[\widthof{$\times$}][c]{0}
\end{pmatrix},
\\
(112)\;\;\;\;\;\;\Sigma_1:\begin{pmatrix}
\times&\makebox[\widthof{$\times$}][c]{0}&\makebox[\widthof{$\times$}][c]{0}
\\
\makebox[\widthof{$\times$}][c]{0} & \makebox[\widthof{$\times$}][c]{0}& \times
\\
\makebox[\widthof{$\times$}][c]{0} & \makebox[\widthof{$\times$}][c]{0}& \makebox[\widthof{$\times$}][c]{0}
\end{pmatrix},
\;\;\Sigma_2:\begin{pmatrix}
\makebox[\widthof{$\times$}][c]{0}&\makebox[\widthof{$\times$}][c]{0}&\makebox[\widthof{$\times$}][c]{0}
\\
\makebox[\widthof{$\times$}][c]{0} & \makebox[\widthof{$\times$}][c]{0}& \makebox[\widthof{$\times$}][c]{0}
\\
\makebox[\widthof{$\times$}][c]{0} & \times& \makebox[\widthof{$\times$}][c]{0}
\end{pmatrix},
\\
(121)\;\;\;\;\;\;\Sigma_1:\begin{pmatrix}
\times&\makebox[\widthof{$\times$}][c]{0}&\makebox[\widthof{$\times$}][c]{0}
\\
\makebox[\widthof{$\times$}][c]{0} & \makebox[\widthof{$\times$}][c]{0}&\makebox[\widthof{$\times$}][c]{0}
\\
\makebox[\widthof{$\times$}][c]{0} &  \times& \makebox[\widthof{$\times$}][c]{0}
\end{pmatrix},
\;\;\Sigma_2:\begin{pmatrix}
\makebox[\widthof{$\times$}][c]{0}&\makebox[\widthof{$\times$}][c]{0}&\makebox[\widthof{$\times$}][c]{0}
\\
\makebox[\widthof{$\times$}][c]{0} & \makebox[\widthof{$\times$}][c]{0}&\times 
\\
\makebox[\widthof{$\times$}][c]{0} & \makebox[\widthof{$\times$}][c]{0}& \makebox[\widthof{$\times$}][c]{0}
\end{pmatrix},
\\
(211)\;\;\;\;\;\;\Sigma_1:\begin{pmatrix}
\makebox[\widthof{$\times$}][c]{0}&\makebox[\widthof{$\times$}][c]{0}&\makebox[\widthof{$\times$}][c]{0}
\\
\makebox[\widthof{$\times$}][c]{0} & \makebox[\widthof{$\times$}][c]{0}&\times 
\\
\makebox[\widthof{$\times$}][c]{0} &  \times& \makebox[\widthof{$\times$}][c]{0}
\end{pmatrix},
\;\;\Sigma_2:\begin{pmatrix}
\times&\makebox[\widthof{$\times$}][c]{0}&\makebox[\widthof{$\times$}][c]{0}
\\
\makebox[\widthof{$\times$}][c]{0} & \makebox[\widthof{$\times$}][c]{0}&\makebox[\widthof{$\times$}][c]{0}
\\
\makebox[\widthof{$\times$}][c]{0} & \makebox[\widthof{$\times$}][c]{0}& \makebox[\widthof{$\times$}][c]{0}
\end{pmatrix},
\end{split}
\end{align*}
and so on for 122, 212, 221, 222. 

\subsection*{Anomaly cancellation conditions}

While global anomalies are typically regarded as benign, one must recall that the work previously developed in \cite{Astrid:2019,Astrid:2020}, where this article is inspired on, considers a local $\U{}^\prime$ symmetry where gauge anomalies must necessarily be forbidden. With this in mind, and with the purpose of making the considered model consistent with a gauged version (to be studied elsewhere), one must also include a set of restrictions that incorporate the $\U{}^\prime$ charges in the following triangle anomalies
\begin{align}
\begin{split}
\left[\mathrm{U(1)^\prime}\right]^3, \;\;\mathrm{U(1)^\prime}\left[\mathrm{Gravity}\right]^2,
\\
\mathrm{U(1)^\prime} \left[\mathrm{U(1)}_\mathrm{Y}\right]^2, \;\;\mathrm{U(1)^\prime}\left[\mathrm{SU(2)_L}\right]^2,
\\
\mathrm{U(1)^\prime}\left[\mathrm{SU(3)_C}\right]^2, \;\; \left[\mathrm{U(1)^\prime}\right]^2\mathrm{U(1)_Y}\,.
\end{split}
\end{align}
As the right-handed neutrinos are solely charged under $\mathrm{U(1)^\prime}$, confer \cref{tab:model}, the only anomalies altered in comparison to \cite{Astrid:2019} are those coming from $\left[\mathrm{U(1)^\prime}\right]^3$ and $\mathrm{U(1)^\prime}\left[\mathrm{Gravity}\right]^2$, \textit{i.e.}
\begin{align}\label{Anomaly_Cond_new}
\begin{split}
A_{_{\mathrm{U(1)^\prime}\mathrm{U(1)^\prime}\mathrm{U(1)^\prime}}}&\equiv \sum_{i=1}^{3}\left( 6X_{q_i}^3 + 2X_{l_i}^3 - 3X_{u_i}^3 - 3X_{d_i}^3 - X_{e_i}^3 - X_{\nu_i}^3\right) = 0,\\ 
A_{_{\mathrm{g}\mathrm{g}\mathrm{U(1)^\prime}}}&\equiv\sum_{i=1}^{3} \left(6X_{q_i} + 2X_{l_i} - 3X_{u_i} - 3X_{d_i} - X_{e_i} - X_{\nu_i}  \right) = 0\,,
\end{split}
\end{align}
where $g$ denotes gravitational interactions while $A_{_{X,Y,Z}} \propto \mathrm{Tr}[ \mathrm{T}_X \{\mathrm{T}_Y,\mathrm{T}_Z\}]$ with $\mathrm{T}$ being the generators of the corresponding Lie Algebra. The remaining anomaly cancellation conditions are given in \cref{Anomaly_Cond}. Overall, this results in a series of equations for anomaly cancellation, restricting the possible values of $X_{\Psi_i}$ from where one can extract all viable, anomaly-free model versions.

\subsection*{Anomaly-free solutions}\label{s:AnomaFree}

Focusing on the solutions compatible with the BGL model, there are in total three anomaly-free possibilities equipped with a type-I seesaw mechanism. These are denoted as $\nu$BGL-I, $\nu$BGL-IIa and $\nu$BGL-IIb, being characterized by their lepton sector textures as follows 
\begin{enumerate}
    \item $\nu$BGL-I Scenario
    \begin{align}\label{nuBGL-I Scenario}
\begin{split}
\Pi_1,\Sigma_1,B=
\begin{pmatrix}
\times&\times& \makebox[\widthof{$\times$}][c]{0}\\
\times&\times& \makebox[\widthof{$\times$}][c]{0}\\
 \makebox[\widthof{$\times$}][c]{0}& \makebox[\widthof{$\times$}][c]{0}& \makebox[\widthof{$\times$}][c]{0}
\end{pmatrix}\,,\quad
\Pi_2,\Sigma_2=
\begin{pmatrix}
 \makebox[\widthof{$\times$}][c]{0}& \makebox[\widthof{$\times$}][c]{0}& \makebox[\widthof{$\times$}][c]{0}\\
 \makebox[\widthof{$\times$}][c]{0}& \makebox[\widthof{$\times$}][c]{0}& \makebox[\widthof{$\times$}][c]{0}\\
 \makebox[\widthof{$\times$}][c]{0}& \makebox[\widthof{$\times$}][c]{0}&\times
\end{pmatrix}\,,\\
A=\mathbb{0}\,,\quad
C=
\begin{pmatrix}
 \makebox[\widthof{$\times$}][c]{0}& \makebox[\widthof{$\times$}][c]{0}&\times\\
 \makebox[\widthof{$\times$}][c]{0}& \makebox[\widthof{$\times$}][c]{0}&\times\\
\times&\times& \makebox[\widthof{$\times$}][c]{0}
\end{pmatrix}\,.
\end{split}
\end{align}
\item $\nu$BGL-IIa Scenario
\begin{align}
\begin{split}
\Pi_1&,\Sigma_1=
\begin{pmatrix}
\times& \makebox[\widthof{$\times$}][c]{0}& \makebox[\widthof{$\times$}][c]{0}\\
 \makebox[\widthof{$\times$}][c]{0}&\times& \makebox[\widthof{$\times$}][c]{0}\\
 \makebox[\widthof{$\times$}][c]{0}& \makebox[\widthof{$\times$}][c]{0}& \makebox[\widthof{$\times$}][c]{0}
\end{pmatrix}\,,\\
\Pi_2&=
\begin{pmatrix}
 \makebox[\widthof{$\times$}][c]{0}& \makebox[\widthof{$\times$}][c]{0}& \makebox[\widthof{$\times$}][c]{0}\\
\times& \makebox[\widthof{$\times$}][c]{0}& \makebox[\widthof{$\times$}][c]{0}\\
 \makebox[\widthof{$\times$}][c]{0}& \makebox[\widthof{$\times$}][c]{0}&\times
\end{pmatrix}\,,\quad
\Sigma_2=
\begin{pmatrix}
 \makebox[\widthof{$\times$}][c]{0}&\times& \makebox[\widthof{$\times$}][c]{0}\\
 \makebox[\widthof{$\times$}][c]{0}& \makebox[\widthof{$\times$}][c]{0}& \makebox[\widthof{$\times$}][c]{0}\\
 \makebox[\widthof{$\times$}][c]{0}& \makebox[\widthof{$\times$}][c]{0}&\times
\end{pmatrix}\, \\
A&=\begin{pmatrix}
\times& \makebox[\widthof{$\times$}][c]{0}& \makebox[\widthof{$\times$}][c]{0}\\
 \makebox[\widthof{$\times$}][c]{0}& \makebox[\widthof{$\times$}][c]{0}& \makebox[\widthof{$\times$}][c]{0}\\
 \makebox[\widthof{$\times$}][c]{0}& \makebox[\widthof{$\times$}][c]{0}& \makebox[\widthof{$\times$}][c]{0}
\end{pmatrix}\,,\quad
B=\begin{pmatrix}
 \makebox[\widthof{$\times$}][c]{0}& \makebox[\widthof{$\times$}][c]{0}& \makebox[\widthof{$\times$}][c]{0}\\
 \makebox[\widthof{$\times$}][c]{0}& \makebox[\widthof{$\times$}][c]{0}&\times\\
 \makebox[\widthof{$\times$}][c]{0}&\times& \makebox[\widthof{$\times$}][c]{0}
\end{pmatrix}\,,\quad
C=
\mathbb{0}
\,.
\end{split}
\end{align}
\item $\nu$BGL-IIb Scenario
\begin{align}
\begin{split}
A=\begin{pmatrix}
 \makebox[\widthof{$\times$}][c]{0}& \makebox[\widthof{$\times$}][c]{0}& \makebox[\widthof{$\times$}][c]{0}\\
 \makebox[\widthof{$\times$}][c]{0}& \makebox[\widthof{$\times$}][c]{0}&\times\\
 \makebox[\widthof{$\times$}][c]{0}&\times& \makebox[\widthof{$\times$}][c]{0}
\end{pmatrix}\,,\quad
B=\begin{pmatrix}
\times& \makebox[\widthof{$\times$}][c]{0}& \makebox[\widthof{$\times$}][c]{0}\\
 \makebox[\widthof{$\times$}][c]{0}& \makebox[\widthof{$\times$}][c]{0}& \makebox[\widthof{$\times$}][c]{0}\\
 \makebox[\widthof{$\times$}][c]{0}& \makebox[\widthof{$\times$}][c]{0}& \makebox[\widthof{$\times$}][c]{0}
\end{pmatrix}\,,\quad
C=\mathbb{0}
\,.
\end{split}
\end{align}
\end{enumerate}
where what distinguishes the $\nu$BGL-IIa and $\nu$BGL-IIb scenarios are the textures in the Majorana neutrino sector. Last but not least we show in \cref{tab:charges} the generic $\U{}^\prime$ charges for each of the BGL scenarios.
\begin{table*}[htb!]
	\centering
	{\small
		\begin{tabular}{l l l cr}
	Charges&$\nu$BGL-I&$\nu$BGL-IIa&$\nu$BGL-IIb&\\
			\hline\hline\\
		$q_L$&$\left[\begin{array}{c}x\\x\\x_{tL}\end{array}\right]$&$--$&$--$\\[1cm]
		$u_R$&$\left[\begin{array}{c}y\\y\\x_{tR}\end{array}\right]$&$--$&$--$\\[1cm]
		$d_R$&$\left[\begin{array}{c}2x-y\\2x-y\\2x-y\end{array}\right]$&$--$&$--$\\[1cm]
		$\ell_L$& $\left[\begin{array}{c}-3x\\-3x\\21x-6y\end{array}\right]$&$\left[\begin{array}{c}x-y\\-7x+y\\21x-6y\end{array}\right]$&$\dfrac{1}{3}\left[\begin{array}{c}-x-2y\\-17x+2y\\39x-12y\end{array}\right]$\\[1cm]
		$e_R$&$\left[\begin{array}{c}-2x-y\\-2x-y\\30x-9y\end{array}\right]$&$\left[\begin{array}{c}2x-2y\\-6x\\30x-9y\end{array}\right]$&$\dfrac{1}{3}\left[\begin{array}{c}2x-5y\\-14x-y\\58x-19y\end{array}\right]$\\[1cm]
		$\nu_R$&$\left[\begin{array}{c}-4x+y\\-4x+y\\12x-3y\end{array}\right]$&$\left[\begin{array}{c}0\\-8x+2y\\12x-3y\end{array}\right]$&$\dfrac{1}{3}\left[\begin{array}{c}-4x+y\\-20x+5y\\20x-5y\end{array}\right]$\\[1cm]
		$\Phi$&$\left[\begin{array}{c}-x+y\\-9x+3y\end{array}\right]$&$\left[\begin{array}{c}-x+y\\-9x+3y\end{array}\right]$&$\dfrac{1}{3}\left[\begin{array}{c}3(-x+y)\\-19x+7y\end{array}\right]$\\[1cm]
		$S$&$8x-2y$&$-4x+y$&$\dfrac{8x-2y}{3}$\\[0.5cm]
		\hline\hline\\
		\end{tabular}
	}
	\caption{Allowed charges for the various models. For model $\nu$BGL-I and IIa we have $x_{tL}=-7x+2y$ and $x_{tR}=-16x+5y$. Model $\nu$BGL-IIb has $x_{tL}=(-13x+4y)/3$ and $x_{tR}=(-32x+11y)/3$. In order for the BGL textures to be preserved, we additionally require that $y\neq 4x$.}
	\label{tab:charges}
\end{table*}

\section{Chosen scenario}\label{sec:chosen_scenario}

In what follows we will focus on the $\nu$BGL-I scenario, whose allowed charges are constrained by the first column in \cref{tab:charges}. We make this choice after having investigated the other possibilities available within
this model and verified that the $\nu$BGL-I scenario is the one for which our numerical analysis and parameter space
scans were more efficient. This, of course, does not exclude the possible interest of the other scenarios, but our main
goal is to understand the typical phenomenology of this class of models. Thus, for $\nu$BGL-I, 
choosing $x=1$ and $y=1/3$ the model considered in this article is then defined by the quantum numbers in \cref{tab:model}.
\begin{table}[thb!]
    \captionsetup{justification=raggedright,singlelinecheck=false}
	\begin{tabular}{|c|ccc|ccc|ccc|ccc|}
		\hline
		& $\Phi_{1}$ & $\Phi_{2}$ & $S$ & $q_{1}$ & $q_{2}$ & $q_{3}$ & $u_{R_1}$ & $u_{R_2}$ & ${u}_{R_3}$ & $d_{R_1}$ & $d_{R_2}$ & $d_{R_3}$ \\ \hline\hline
		$\mathrm{U(1)_Y}$ & 1/2 & 1/2 & 0 & 1/6 & 1/6 & 1/6 & 2/3 & 2/3 & 2/3 & $-$1/3 & $-$1/3& $-$1/3\\
		$\mathrm{SU(2)_L}$ & \bf{2}& \bf{2}& \bf{1}& \bf{2} & \bf{2} & \bf{2} & \bf{1} & \bf{1} & \bf{1} & \bf{1} & \bf{1}& \bf{1}\\
        $\mathrm{SU(3)_C}$& \bf{1}& \bf{1}& \bf{1} & \bf{3} & \bf{3} & \bf{3} & \bf{3} & \bf{3} & \bf{3} & \bf{3} & \bf{3}& \bf{3}\\
		$\mathrm{U(1)^{\prime}}$ &$-$2/3 & $-$8 & 22/3 & 1 & 1 & $-$19/3 & 1/3 & 1/3 & $-$43/3 & 5/3 & 5/3& 5/3\\ \hline
	\end{tabular} \\[0.5 cm]
	\begin{tabular}{|c|ccc|ccc|ccc|}
		\hline
		& $\ell_{1}$ & $\ell_{2}$ & $\ell_{3}$ & $e_{R_1}$ & $e_{R_2}$ & $e_{R_3}$ & $\nu_{R_1}$ & $\nu_{R_2}$ & $\nu_{R_3}$ \\ \hline\hline
		$\mathrm{U(1)_Y}$ & $-$1/2 & $-$1/2 & $-$1/2 & $-$1 & $-$1 & $-$1 & 0 & 0 & 0 \\
		$\mathrm{SU(2)_L}$ & \bf{2} & \bf{2} & \bf{2} & \bf{1} & \bf{1} & \bf{1} & \bf{1} & \bf{1}& \bf{1}\\
        $\mathrm{SU(3)_C}$ & \bf{1} & \bf{1} & \bf{1} & \bf{1} & \bf{1} & \bf{1} & \bf{1} & \bf{1}& \bf{1}\\
		$\mathrm{U(1)^{\prime}}$ & $-$3 & $-$3 & 19 & $-$7/3 & $-$7/3 & 27 & $-$11/3 & $-$11/3 & 11 \\ \hline
	\end{tabular}%
	\caption{Charge assignments of the SU(2) scalar doublets and SM fermions
		under the flavour symmetry $\mathrm{U(1)^{\prime}}$ and the SM gauge group, $\mathrm{SU(3)_C\times SU(2)_L\times U(1)_Y}$. The corresponding $\mathrm{U(1)'}$ charges are computed using the formulas from the first column of \cref{tab:charges} taking $x=1$ and $y=1/3$.}
	\label{tab:model}
\end{table}
\\
Notice that the assigned quantum numbers for the up-type quarks can also be flipped in such a way that there are in total three variations of the $\nu$BGL-I scenario. This depends, in practice, on the choice of up-type quark associated with the $\Delta_{2}$ matrix in Eq.~\eqref{BGLtextures} and the designation of the model is typically associated with that, \textit{i.e.}~variation $u$, $c$ or $t$. In the reminder of this article we focus on variation $t$. 

As already stated, one of the novel features of the  model considered is the presence of three generations of right-handed neutrinos. In order to comply with experimental data a realistic lepton mixing and neutrino mass hierarchy must be assured. Both the charged neutral sectors can be rotated to the mass basis via bi-unitary matrices, that is,
\begin{equation}\label{eq:biunitary_trans}
\begin{aligned}
&M_\mathrm{L}^{\mathrm{diag}} = U^\dagger_\mathrm{L} M_\mathrm{L} U_\mathrm{R}, \\
&M_\mathrm{\nu}^{\mathrm{diag}} = U^\dagger_\nu M_\nu U_\nu,
\end{aligned}
\end{equation}
where $M_\mathrm{L}$ is a $3\times 3$ matrix expressed in the basis $\{e_\mathrm{L}^{1}, e_\mathrm{L}^{2}, e_\mathrm{L}^{3}\} \otimes \{ e_\mathrm{R}^{1}, e_\mathrm{R}^{2}, e_\mathrm{R}^{3} \}$ as usual, while $M_\mathrm{\nu}$ is a $6\times 6$ matrix written in the basis $\{\nu_\mathrm{L}^{1}, \nu_\mathrm{L}^{2}, \nu_\mathrm{L}^{3}, \nu_\mathrm{R}^{1}, \nu_\mathrm{R}^{2}, \nu_\mathrm{R}^{3}\}$. On the other hand, $U_\mathrm{L}$, $U_\mathrm{R}$ and $U_\nu$ are unitary matrices. We can then define the PMNS matrix as
\begin{equation}\label{eq:PMNS_matrix}
V_P = U^\dagger_\mathrm{L} P U_\nu = \begin{pmatrix}
V_{\mathrm{PMNS}}^{\mathrm{SM}} & \mathcal{O}_{P}
\end{pmatrix}, \quad P = \begin{pmatrix}
\mathbb{1}_{3\times 3} \hspace{0.5em} | \hspace{0.5em} 0_{3\times 3}
\end{pmatrix},
\end{equation}
where $V_{P}$ is a $3\times 6$ matrix with the first $3\times3$ entries  encoding the SM charged-lepton and neutrinos interactions and $\mathcal{O}_{P}$ contains the interactions between leptons and the heavy  neutrinos. In our analysis we require $\mathcal{O}_{P} < 10^{-5}$ such that charged currents involving active and Beyond Standard Model (BSM) neutrinos are suppressed. Also, we require the masses of such active neutrinos to be between [0,$10^{-9}$]~GeV while their heavy counterparts have masses around the TeV scale. While an in-depth study of neutrino physics is beyond the scope of this work, we have performed a fit to the light neutrinos masses in order to respect the normal hierarchy \cite{Zyla:2020zbs},
\begin{equation}\begin{split}
   M_{\nu_2}^2 - M_{\nu_1}^2 \approx 10^{-5}~\mathrm{eV^2},\\
   M_{\nu_3}^2 - M_{\nu_2}^2 \approx 10^{-3}~\mathrm{eV^2}.
   \end{split}
\end{equation}
This was numerically required for all points that pass the theoretical and experimental constraints, as it will be discussed in upcoming sections. Although a full fit was not done across the entirety of the parameter space, and given  that our studies of the scalar and quark sectors are independent of neutrino physics, we have confirmed and can safely argue that the discussed model is compatible with neutrino data.

\subsection*{Inversion procedure for the scalar sector}

Following the same methodology used previously in \cite{Das:2021oik}, our goal is to provide the physical masses and mixing angles as inputs and then search for valid regions in the parameter space. In the considered $\nu$BGL-I scenario the scalar potential is given in \cref{Scal_Pote_1} with $a_3,a_4 = 0$. Selecting a vacuum configuration with real VEVs $(v_1,v_2,v_S)$, one obtains the following minimization conditions
\begin{align}
\label{Scal_Pote_2}
\begin{split}
&\mu_1^2 =\quad - \left(v_1^2\lambda_1 + \frac{v_S^2 \lambda^\prime_2}{2} +\frac{v_2\hat{\mu}}{2v_1} \right)\:,\\
&\mu_2^2 =\quad - \left(v_2^2\lambda_2 +\frac{v_S^2\lambda^\prime_3}{2} +\frac{v_1\hat{\mu}}{2v_2} \right)\:,\\
&\mu_S^2 =\quad -\frac{\sqrt{2}v_1 v_2\left(a_1 +a_2\right)+v_S\left( v_1^2 \lambda_2^\prime + v_2^2\lambda^\prime_3 +2v_S^2\lambda^\prime_1 + 2\mu_{b}^2\right) }{2v_S}\,,
\end{split}
\end{align}
where one defines
\begin{align}
\label{Scal_Pote_3}
\begin{split}
\hat{\mu} = v_S \sqrt{2}\left(a_1+a_2\right) + v_1 v_2\left(\lambda_3 + \lambda_4\right)+2\mu_3^2\,.
\end{split}
\end{align}
The physical scalar spectrum contains three CP-even scalars $H_{1,2,3}$, two pseudo-scalars $A_{2,3}$, and one charged Higgs boson $H^\pm$. In a configuration close to the vacuum alignment in one of the Higgs doublets, the corresponding CP-even scalar corresponds to the SM Higgs boson with mass $125.09~\mathrm{GeV}$ and with SM-like interactions to the gauge bosons and fermions. Notice that, if such interactions coincide with the SM ones, one often refers to it as the exact alignment limit. However, direct searches for new scalars \cite{ATLAS:2018sbw,CMS:2015uzk,CMS:2016xnc,ATLAS:2019tpq,ATLAS:2020tlo,CMS:2019ogx} still keep a window opened for possible off-alignment deviations that we consider in our analysis below. In order to keep the physical parameters of the theory under control one performs an inversion procedure analogous to that developed in \cite{Das:2021oik}, expressing the parameters of the scalar potential in \cref{Scal_Pote_2} in terms of physical masses, mixing angles (including the Higgs alignment parameter) and VEVs as we discuss in what follows.

First, the matrix $\mathcal{O}_{\beta}$, which rotates the gauge eigenstates into the Higgs basis\footnote{The Higgs basis corresponds to a redefinition of the physically-equivalent doublets so that only one of them contains a VEV.}, is introduced
\begin{align}
\label{Obeta}
\begin{split}
\mathcal{O}_\beta = \begin{pmatrix}
 \cos\beta & \sin\beta & 0 \\
 -\sin\beta & \cos\beta & 0 \\
 0 & 0 & 1
\end{pmatrix} = \begin{pmatrix}
 v_1/v & v_2/v & 0 \\
 - v_2/v& v_1/v & 0 \\
 0 & 0 & 1
\end{pmatrix}
\end{split}
\end{align}
where one defines
\begin{align}\begin{split}
    &v_1  = v\cos\beta\,,\\
    &v_2  = v\sin\beta\,,\\
    &v =\sqrt{{v_1}^2+{v_2}^2} \approx 246\ \mathrm{GeV}\, .
\end{split}\end{align}
Starting with the scalar potential in \cref{eq:Scalar_pot} and expanding it using the field definitions in \cref{eq:Field_ext}, the CP-odd sector mass matrix can be obtained as
\begin{align}\label{eq:mathcal_Ms}
\frac{1}{2}\begin{pmatrix}
I_1 & I_2 & \eta\end{pmatrix}{\mathcal{M}_S}^2\begin{pmatrix}
     I_1 \\ I_2 \\ \eta\end{pmatrix}\,.
\end{align}
Using $\mathcal{O}_\beta$ one can rotate $\mathcal{M}_S^2$ to the Higgs basis, where it acquires a block-diagonal form 
\begin{align}\label{eq:mathbb_MS}
    {\mathfrak{M}_S}^2 = \mathcal{O}_{\beta} \mathcal{M}_S^2 \mathcal{O}_{\beta}^{T} \, .
\end{align}
We can further diagonalize ${\mathfrak{M}_S}^2 $ via an additional rotation matrix $\mathcal{O}_\gamma$ as
\begin{align}\label{eq:MA mass}
\begin{pmatrix}
 0 & 0 & 0\\
 0 & {M_{A_2}}^2 & 0 \\
 0 & 0 & {M_{A_3}}^2 
\end{pmatrix} = \mathcal{O}_\gamma{\mathfrak{M}_S}^2{\mathcal{O}_\gamma}^T =  \mathcal{O}_\gamma\mathcal{O}_{\beta} \mathcal{M}_S^2 \mathcal{O}_{\beta}^{T}{\mathcal{O}_\gamma}^T
\end{align}
where one defines
\begin{align}\label{eq:gamma}
\mathcal{O}_\gamma = \begin{pmatrix}
 1 & 0 & 0\\
 0 & \cos\gamma_1 & - \sin\gamma_1 \\
 0 & \sin\gamma_1 & \cos\gamma_1 
\end{pmatrix} \,.
\end{align}
Notice that while the neutral CP-odd Goldstone boson can be gauged away from the particle spectrum and gets absorbed as the longitudinal polarisation of the $\mathrm{Z^0}$ boson, one of the two physical pseudo-scalars is in fact a pseudo-Goldstone state that emerges due to both spontaneous and explicit breaking of the global $\U{}^\prime$ symmetry.

Similarly, the CP-even states can be rotated to the mass basis as
\begin{align}\label{eq:M_hmatrix_eq_1}
\begin{pmatrix}
{M_{H_1}}^2 & 0 & 0\\
 0 & {M_{H_2}}^2 & 0 \\
 0 & 0 & {M_{H_3}}^2 
\end{pmatrix} =  \mathcal{O}_\alpha\mathcal{O}_{\beta} \mathcal{M}_H^2 \mathcal{O}_{\beta}^{T}{\mathcal{O}_\alpha}^T
\end{align}
with 
\begin{align}
    \mathcal{O}_{\alpha} = R_3\dotproduct  R_2\dotproduct R_1 
\end{align}
and
\begin{align}\label{eq:M_hmatrix_eq_2}\begin{split}
    R_1 = \begin{pmatrix}
     \cos{\alpha_1} & -\sin{\alpha_1} & 0\\ \sin{\alpha_1} & \cos{\alpha_1} & 0\\ 0 & 0 & 1\\\end{pmatrix},
     R_2 = \begin{pmatrix}
     \cos{\alpha_2} & 0 & \sin{\alpha_2}\\ 0 & 1 & 0\\
     -\sin{\alpha_2} & 0 & \cos{\alpha_2}\\\end{pmatrix}, R_3 = \begin{pmatrix} 1 & 0 & 0\\
     0 & \cos{\alpha_3} & \sin{\alpha_3}\\ 0 & -\sin{\alpha_3} & \cos{\alpha_3} \end{pmatrix}\,.
\end{split}\end{align}
The presence of a CP-even state behaving as a SM-like Higgs boson is ensured not only requiring that
\begin{align}
    M_{H_1}= 125.09\ \textrm{GeV}\ ,
\end{align}
but also that the couplings to fermions and gauge bosons of this particle closely reproduce the 
SM predictions, which are controlled by the coupling modifier
\begin{equation}
    \left[ \mathcal{O}_\alpha\mathcal{O}_{\beta} \right]_{12} = - \cos \alpha_2 \sin \delta \qquad \textrm{with} \qquad \delta \equiv \alpha_1 - \beta
    \label{eq:alignment}
\end{equation}
such that the coupling of the 125 GeV Higgs boson to the electroweak gauge bosons can be expressed as
\begin{align}
    \lambda^\prime_{h\,V\,V} = - \lambda^{SM}_{h\,V\,V} \cos \alpha_2 \sin \delta
    \label{eq:hSMSM}
\end{align}
with $\lambda^{SM}_{h\,V\,V}$ denoting the SM Higgs boson coupling to gauge bosons, $V \,=\, Z,\, W$ in the SM. 
This means that the alignment condition concerning Higgs gauge interactions is governed not by a single angle but instead two angular parameters, $\delta$ and $\alpha_2$. Furthermore, as the LHC has been putting stringent limits on deviations from the SM, one expects that $\cos \alpha_2 \sin \delta \sim \mathcal{O}(1)$, \textit{i.e.}~$\delta$ must not be far from $\pi/2$ while $\alpha_2$ must be small. Note also that Eq.~\eqref{eq:alignment} ensures that the interactions of the SM-like Higgs boson are dominantly inherited from the second doublet %expressed in the Higgs basis, 
which is the one that couples to the top-quark. Further conditions for alignment will be obtained from the Higgs' 
interactions with fermions. For instance, given that the top mass in this model comes only from the matrix $\Delta_2$, 
it will be proportional to $v_2$ and therefore the coupling modifier for the Higgs-top interations will be
\begin{align}
    \lambda^\prime_{h\,t\,\bar{t}} = - \lambda^{SM}_{h\,t\,\bar{t}} \frac{\cos \alpha_2 \sin \delta}{\sin\beta}\,.
    \label{eq:htt}
\end{align}
With $\sin\delta = \sin(\alpha_1 - \beta) \sim \mathcal{O}(1)$ as mentioned earlier, we see that a further condition
for alignment are $\cos\alpha_2 \simeq - \sin\beta \simeq -\cos\alpha_1$. Additional conditions would be 
obtained from analysing the 
remaining Higgs-fermion interactions. Notice that the minus signs present in Eqs.~\eqref{eq:alignment}
and~\eqref{eq:htt} are not necessarily meaningful. We know, from the measured Higgs diphoton branching ratio at the
LHC, that the relative sign of $\lambda^\prime_{h\,t\,\bar{t}}$ and $\lambda^\prime_{h\,W\,W}$ ought to be the
same as that of $\lambda^{SM}_{h\,t\,\bar{t}}$ and $\lambda^{SM}_{h\,W\,W}$. An overall minus sign affecting both couplings can therefore be physically meaningless. In our numerical analysis of the next section, however, we
took into account all possible relative signs, and compared the predicted Higgs branching ratios and 
production cross sections with the known experimental values, thus ensuring that, whatever relative signs there
may be between Higgs couplings in this model, the LHC constraints are satisfied.

Last but not least, the charged Higgs sector contains only one physical scalar as the charged Goldstone boson is gauged away from the particle spectrum and plays a role of the longitudinal polarization of the $\mathrm{W}^\pm$ boson. In particular, the mass form reads as
\begin{align}\label{eq:M_Hm_matrix_eq_1}
\begin{pmatrix}
0 & 0\\
 0 & {M_{H^{\pm}}}^2 \\
\end{pmatrix} =  {\mathcal{O}_\gamma}^{\prime}{\mathcal{O}_{\beta}^{\prime}} \mathcal{M}^{\pm}{\mathcal{O}_{\beta}^{\prime}}^{T} {{\mathcal{O}_\gamma}^{\prime}}^{T}
\end{align}
with
\begin{align}
    \mathcal{O}_{\beta}^{\prime}= \begin{pmatrix}
 \cos\beta & \sin\beta\\
 -\sin\beta & \cos\beta
\end{pmatrix}\ , \  \mathcal{O}_{\gamma}^{\prime}= \begin{pmatrix}
 \cos\gamma_2 & \sin\gamma_2\\
 -\sin\gamma_2 & \cos\gamma_2
\end{pmatrix} \, .
\end{align}

The invertion procedure described above is applied to the scalar sector with the purpose of trading the quartic couplings for physical masses and mixing angles. The corresponding analytical formulas are lengthy and can be found in Appendix~\ref{app:analytical_equations}. In our numerical analysis, the physical masses $M_{H_1},M_{H_2},M_{H_3},M_{A_2},M_{A_3},M_{H^{\pm}}$, the angles $\beta,\alpha_1,\alpha_2,\alpha_3,\gamma_1,\gamma_2$ and the cubic couplings $a_1,a_2,a_3$ are then used as input parameters limited to be randomly sampled within physically reasonable ranges as detailed below in Sec.~\ref{sec:Pheno_section}.

\section{Parameter space and phenomenological constraints}\label{sec:Pheno_section}

The parameter space scan discussed in this section relies on the inversion procedure described previously, 
where both scalar masses and mixing angles are used as input, allowing one to determine the gauge eigenbasis couplings of the Higgs potential in \cref{Scal_Pote_1}. The ranges used in the scan can be found in \cref{tab:couplings}.
\begin{table}[htb!]    
	\centering
	\captionsetup{justification=raggedright,singlelinecheck=false}
	\resizebox{.6\textwidth}{!}{\begin{tabular}{c|c|c|c|c}
Parameter & $\alpha_2$,$\alpha_3$,$\gamma_1$ & $\tan \beta$ & $\delta$ & $a_3,a_4
$ \\
\hline
range & $[-\pi,\pi]$ & $[0.5,30]$ & $[\tfrac{\pi}{2}-1,\tfrac{\pi}{2}+1]$ & $0 $ 
\end{tabular}}\\[0.6CM]
	\resizebox{.6\textwidth}{!}{\begin{tabular}{c|c|c|c|c}
Parameter &  $M_{A_2}$, $M_{H^{\pm}}$ & $M_{A_3}$ & $M_{H_2}$,$M_{H_3}$  & $a_2$ \\
\hline
range [GeV] & $[0.5, 1600]$ & $[30, 2000]$  & $[126,1800]$ & $[-1,1]$
\end{tabular}}
	\caption{Parameter intervals used in numerical scans.}
	\label{tab:couplings}
\end{table}
According to the discussion below \cref{eq:hSMSM}, the chosen scanning ranges for $\delta$ are centered around $\pi/2$ where $\sin \delta = 1$. While $\alpha_2$ must be small, we let it range between $-\pi$ and $\pi$ in order to let $\cos \alpha_2$ capture both positive and negative values. Furthermore, the somewhat large ranges set for both $\delta$ and $\alpha_2$ were deliberately picked in order to ensure that the allowed off-alignment regions are entirely covered in the numerical analysis, as we discuss further ahead. It is also worth mentioning that the SM-like Higgs boson was chosen to be the lightest of the CP-even scalars and the absolute value of the cubic couplings is such that $\abs{a_{2}} \leq 1$.

%so that their role in the explicit breakdown of the $\U{}^\prime$ BGL symmetry is sub-leading.

\begin{figure}[htb!]
    \centering
    \vspace{0cm}
    \captionsetup{justification=raggedright,singlelinecheck=false}
	\includegraphics[width=\textwidth]{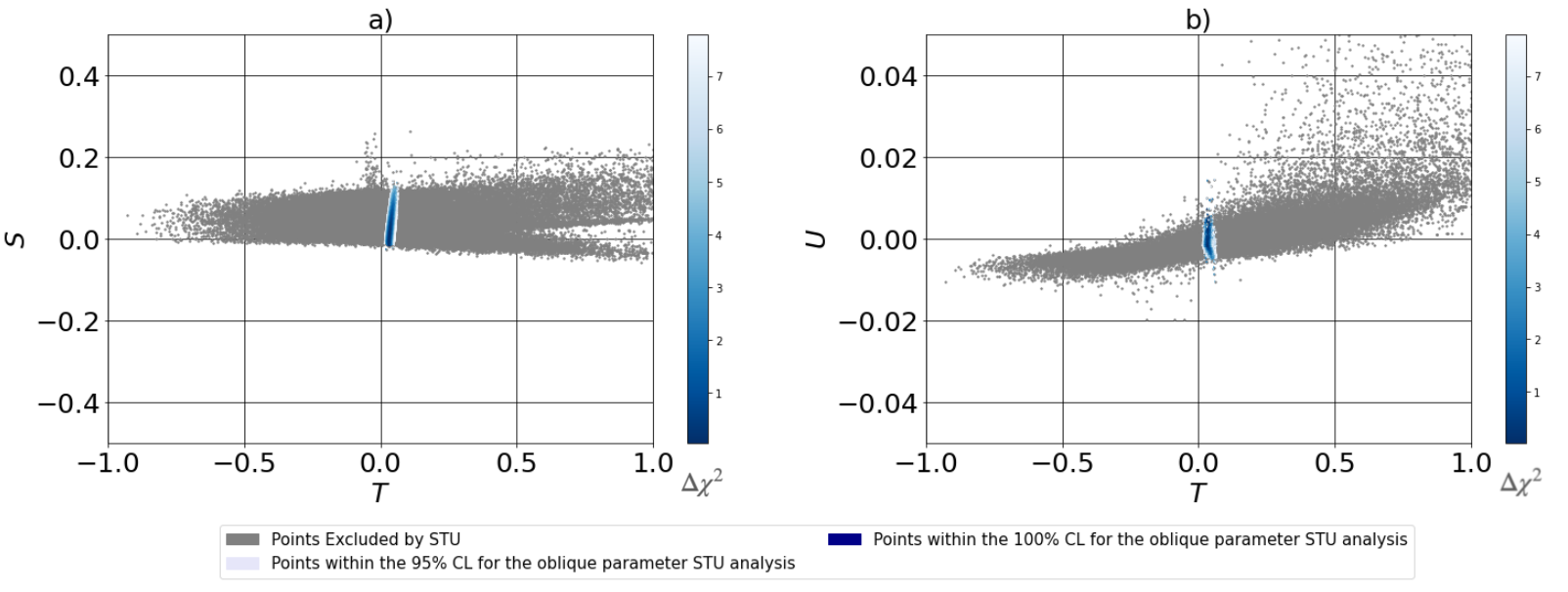}
	\caption{Electroweak precision observables for all simulated points. The colored points are those who pass the STU analysis with a confidence level (CL) of at least 95\%. Grey points are excluded by precision EW fit data.}
	\label{fig:STU95CL}
\end{figure}

\begin{figure}[htb!]
    \centering
    \vspace{0cm}
    \captionsetup{justification=raggedright,singlelinecheck=false}
	\includegraphics[width=\textwidth]{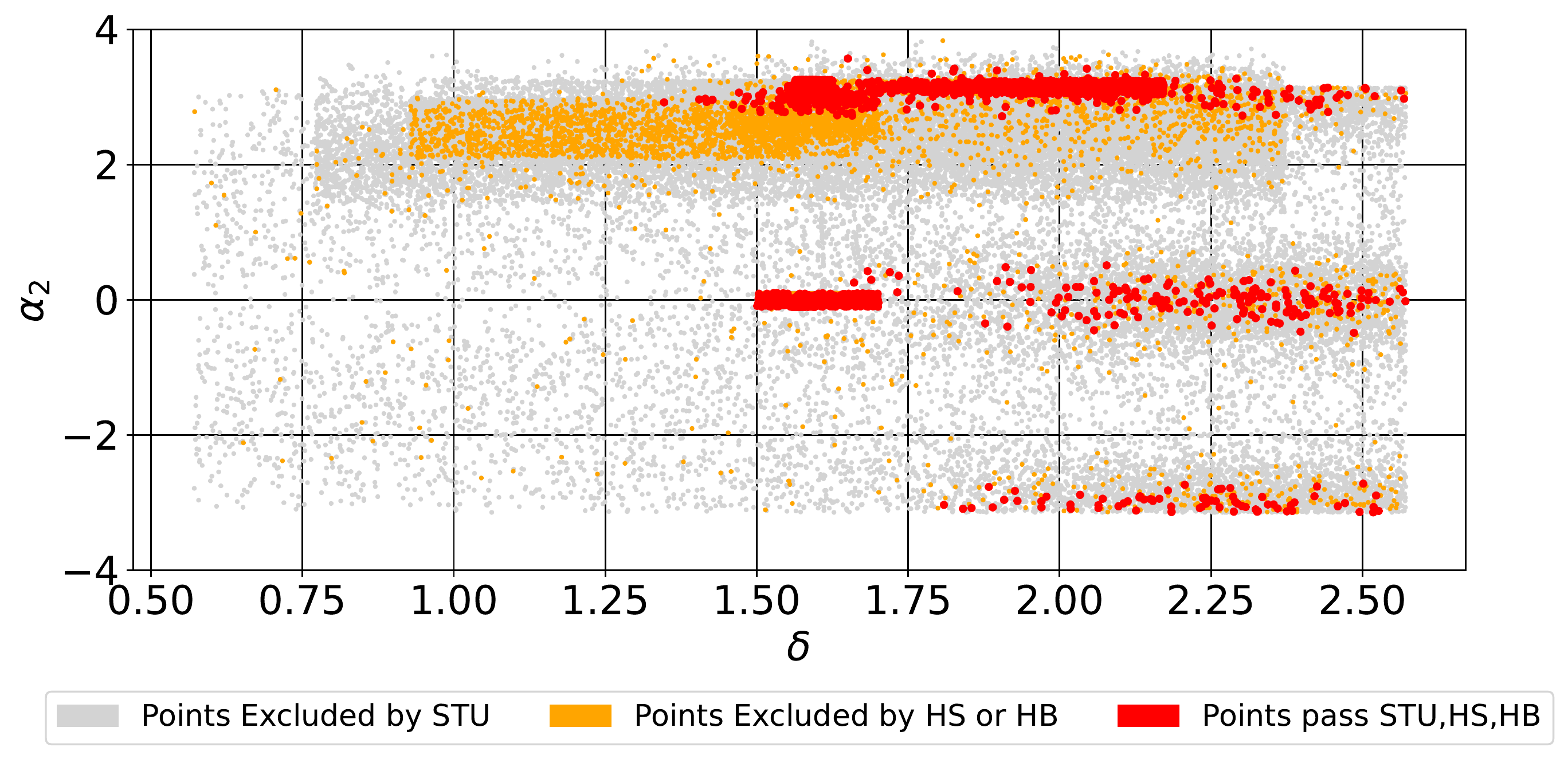}
	\caption{Angles entering in the Higgs coupling modifier of \cref{eq:hSMSM}. Grey points are excluded by EW precision observables while orange ones pass the STU $\Delta \chi^2$ test but fail Higgs physics constraints. Red points pass all restrictions and represent the regions where the Higgs boson is closely aligned with a SM-like one.}
	\label{fig:align}
\end{figure}

Using the inversion procedure we then extract the Lagrangian parameters and feed them to \texttt{SPheno} \cite{Porod:2011nf} in order to calculate both the STU electroweak precision observables (or oblique parameters) as well as Higgs decay widths and branching fractions. The most up to date electroweak fit for the STU parameter \cite{ParticleDataGroup:2020ssz} can be summarized as 
\begin{equation}
\begin{aligned}\begin{split}
\begin{matrix} S& = -0.01 \pm 0.10 \\ T& = 0.03 \pm 0.12 \\  U& = 0.02 \pm 0.11 \end{matrix}\ ,\  \quad\rho_{ij} = \begin{pmatrix} 1 & 0.92 & -0.80\\ 0.92 & 1 & -0.93 \\ -0.80 & -0.93 & 1\\\end{pmatrix}
\end{split}\end{aligned}
\label{STU}
\end{equation}
with $\rho_{ij}$ denoting a $92\%$ correlation between S and T, while U-S and U-T are $80\%$ and $-93\%$ anti-correlated, respectively. For the purpose of this analysis, we required $\Delta\chi^2<7.815$, which translates into a 95\% confidence level (CL) agreement with the electroweak fit, where
\begin{equation}\label{eq:Chi_fit}
\begin{aligned}
\Delta\chi^2 = \sum_{ij}\left(\Delta\mathcal{O}_{i}-\Delta\mathcal{O}_{i}^{(0)}\right)\left[(\sigma^2)^{-1} \right]_{ij}\left(\Delta\mathcal{O}_{j}-\Delta\mathcal{O}_{j}^{(0)}\right)
\end{aligned}
\end{equation}
with $\left[\sigma^2\right]_{ij}\equiv\sigma_i\rho_{ij}\sigma_j$ the covariance matrix and $\sigma_i$ the standard deviations in \cref{STU}. The notation employed here is such that $\Delta\mathcal{O}_{i}^{(0)}$ indicates the $S,T,U$ experimental fit values, while $\Delta\mathcal{O}_{i}$ denotes the $S,T,U$ calculated values in \texttt{SPheno}. In \cref{fig:STU95CL} we show the points that pass the STU restrictions with a CL of, at least, 95\% (blue shades) in comparison with the points that are excluded (grey). The colour bar represents the $\Delta \chi^2$ values. The majority of points that were simulated are rejected due to large $T$ values, as anticipated. However, one of the benefits of 2HDM-like models is that there are parameter space regions for which all constraints on the oblique parameters are satisfied, as is the case here.
While the STU analysis offers a strong razor of the parameter space, it is still necessary to match these points to both Higgs and flavour observables. First, let us focus on the impact of Higgs physics leaving the flavour analysis for later. In \cref{fig:align}, we show the allowed regions in the $\delta$-$\alpha_2$ plane that survive not only the EW precision tests but also that comply with the properties of a SM-like Higgs boson and are not excluded by direct searches for new scalars at the LHC. As usual, this is done by using the \texttt{SPheno} output SLHA files as input cards to \texttt{HiggsSignals} (HS) and \texttt{HiggsBounds} (HB) \cite{Bechtle:2015pma,Bechtle:2013xfa} for every single point that passes the STU test.

\begin{figure}[htb!]
    \centering
    \vspace{0cm}
    \captionsetup{justification=raggedright,singlelinecheck=false}
	\includegraphics[width=\textwidth]{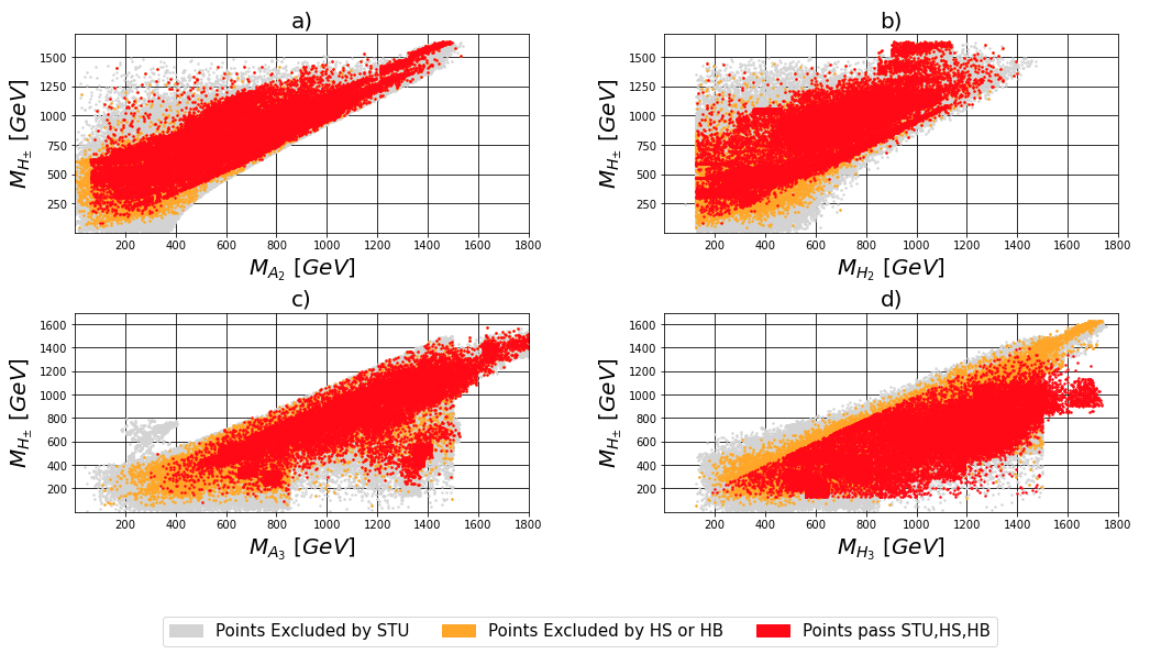}
	\caption{Mass of the charged Higgs versus the masses of the pseudoscalars $A_2$ and $A_3$ (left panels) and of the CP-even scalars $H_2$ and $H_3$ (right panels). Grey points are excluded by the oblique STU parameters, orange points are excluded either by HB or HS,  while simultaneously passing the STU restrictions. Red points pass all constraints.}
	\label{fig:MassP1}
\end{figure}

The points that survive both HS and HB are shown in red and were found to pile in a region around that of the exact alignment limit where $\delta = \pi/2$ and $\alpha_2 = -\pi \,, 0\,, \pi$. In particular, we found that deviations from an exact alignment are primarily driven by the $\delta$ angle which can vary in a wider range, whereas $\alpha_2$ is tightly constrained in the aforementioned regions, with $0.60 < -\cos \alpha_2 \sin \delta < 1.0$ for the upper red island, $0.60 < \cos \alpha_2 \sin \delta < 0.99$ in the middle region, and $0.60 < -\cos \alpha_2 \sin \delta < 0.98$ at the $\alpha_2=-\pi$ line. In \cref{fig:MassP1} we show the dependence of the charged Higgs boson mass $M_{H^\pm}$ in terms of the masses of the $A_2$ and $A_3$ pseudoscalars (left panels) as well as of the masses of the non-SM CP-even states $H_2$ and $H_3$ (right panels). The colour scheme is the same as in \cref{fig:align} and, as the red region demonstrates, we have obtained valid points that cover a wide range of masses for new scalars. In particular, there are viable solutions with charged Higgs bosons that can be as light as approximately $100~\mathrm{GeV}$ extending all the way up to about $1.6~\mathrm{TeV}$, the light pseudo-scalar is limited to the range that goes from $100~\mathrm{GeV}$ up to $1.5~\mathrm{TeV}$ while its heavy counterpart is heavier than $300~\mathrm{GeV}$ but lighter than $1.8~\mathrm{TeV}$. For the CP-even sector, besides a SM-like state with mass $125.09~\mathrm{GeV}$, one finds that $H_2$ is sharply limited from below at the Higgs boson mass, \textit{i.e.}~$125.09~\mathrm{GeV}$, allowing valid solutions up to approximately $1.4~\mathrm{TeV}$, while its heavy partner mass ranges between $200~\mathrm{GeV}$ and approximately $1.7~\mathrm{TeV}$. Note that the limits above must be understood in the context of a search of the parameter space dedicated to find light scalars. In fact, the low mass region is particularly interesting from the collider physics perspective and can potentially be tested in the near future.

\begin{figure}[htb!]
    \centering
    	\scalebox{.3}{\includegraphics{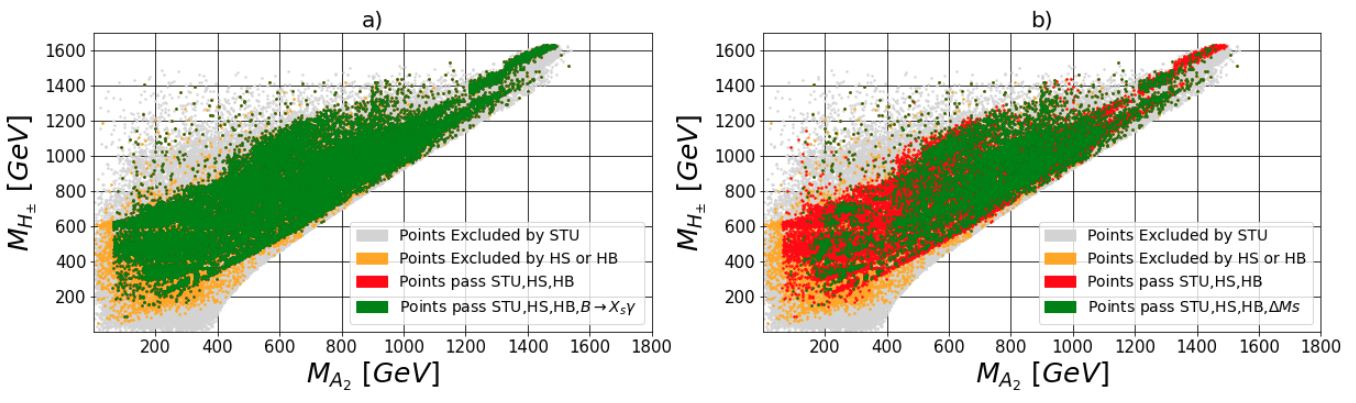}}
        \scalebox{.3}{\includegraphics{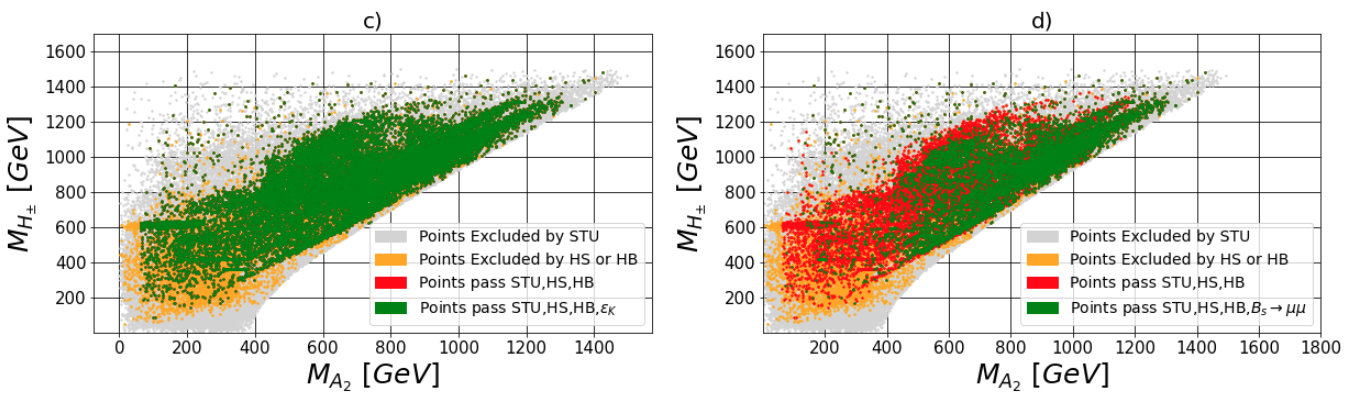}}
        \scalebox{.3}{\includegraphics{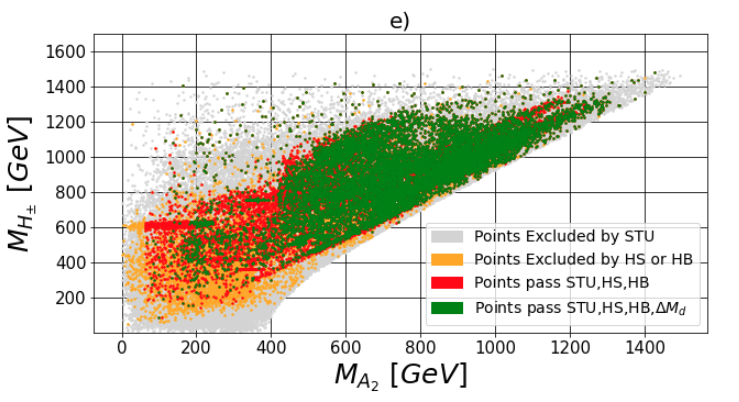}}
    \captionsetup{justification=raggedright,singlelinecheck=false}
    \caption{Mass of the charged Higgs versus the mass of the lightest pseudoscalar $A_2$. Grey points are excluded by STU observables, orange points are excluded by HS or HB while still passing STU, and red points pass STU, HS and HB constraints. In green, we showcase the points that pass a given QFV observable, namely, we have a) $B\rightarrow \chi_S\gamma$, b) $\Delta M_S$, c) $\epsilon_K$, d) $B_S\rightarrow \mu\mu$ and e) $\Delta M_d$, while passing STU, HB and HS constraints.}
    \label{fig:my_label}
\end{figure}

\begin{figure}[htb!]
    \centering
    \captionsetup{justification=raggedright,singlelinecheck=false}
    \scalebox{0.3}{\includegraphics{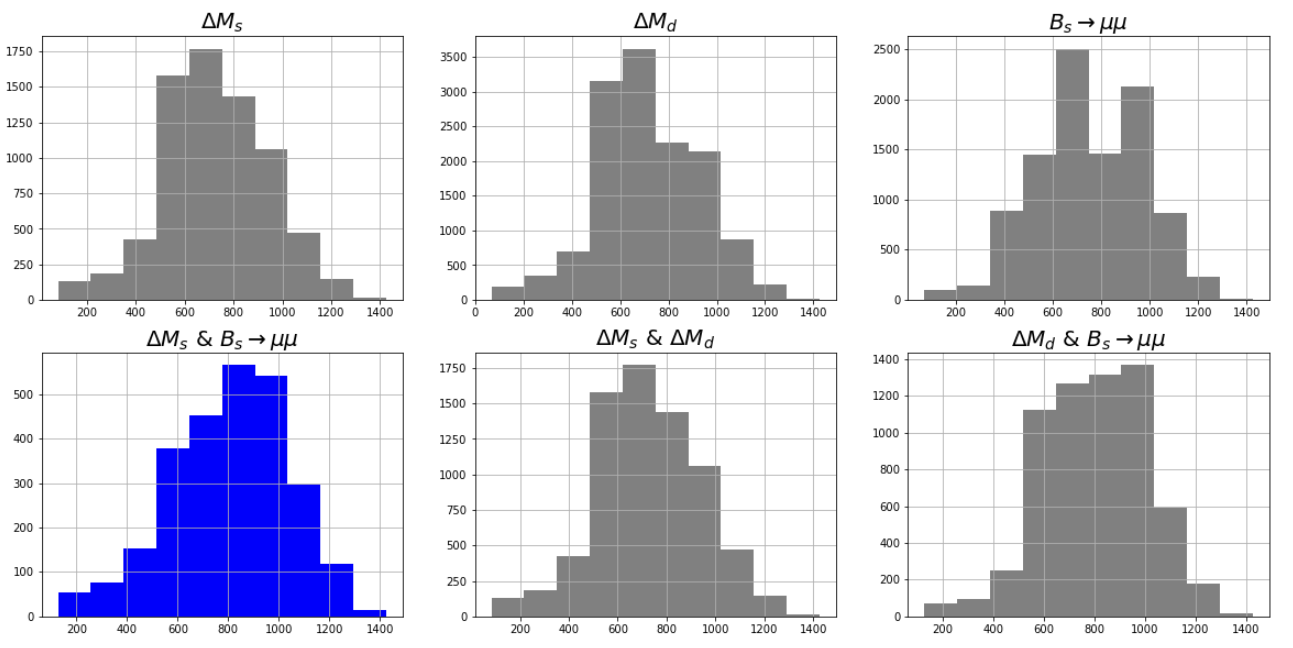}}
    \caption{Histograms containing points that survive STU, HS, HB and a given QFV (or pair of) in bins of the $A_2$ mass. The most restrictive is coloured in {\color{blue} blue}.}
    \label{fig:my_label1111}
\end{figure}

Besides electroweak and Higgs physics constraints, one also has to confront the generated data with quark flavour violation (QFV) observables. According to the textures of the chosen $\nu$BGL-I scenario, the most stringent constraints are on the down-type quark sector rather than in the up-quark or charged lepton ones. Notice also that the most stringent QFV restrictions are indeed coming from processes involving down-type quarks, in particular those shown in \cref{tab:QFV}.
\begin{table}[htb!]    
	\centering
	\captionsetup{justification=raggedright,singlelinecheck=false}
	\resizebox{0.8\textwidth}{!}{\begin{tabular}{c|c c c c c}
Channel & $\mathcal{O}_{SM}$ & $\sigma_{SM}$ & $\mathcal{O}_{Exp}$ & $\sigma_{Exp}$ & $\sigma$ \\[0.1CM] \hline
$\mathrm{BR(B \rightarrow \chi_s \gamma)}$ & $3.29\times10^{-4}$ & $1.87\times 10^{-5}$ & $3.32\times10^{-4}$ & $0.16\times10^{-4}$ & 0.075 \\[0.05CM]
$\mathrm{BR(B_s \rightarrow \mu \mu)}$ & $3.66\times10^{-9}$ & $1.66\times10^{-10}$ &  $2.80\times10^{-9}$ & $0.06\times10^{-9}$ & 0.038 \\[0.05CM]
$\Delta M_d$ (GeV) & $3.97\times10^{-13}$ & $5.07\times10^{-14}$ & $3.33\times10^{-13}$ & $0.013\times10^{-13}$ & 0.11 \\[0.05CM]
$\Delta M_s$ (GeV) & $1.24\times10^{-11}$ & $7.08\times10^{-13}$ & $1.17\times10^{-11}$ & $0.0014\times10^{-11}$ & 0.054 \\[0.05CM]
$\epsilon_K$ (GeV) &$1.81\times10^{-3}$ & $2.00\times10^{-4}$ & $2.23\times10^{-3}$ & $0.011\times10^{-3}$ & 0.14
\end{tabular}}
	\caption{Most relevant Quark Flavour Violation (QFV) observables.
	%Values of QFV as are calculated in \cite{Das:2021oik}. 
	The first column represents the prediction of the SM for each observable, the second column indicates the 1$\sigma$ error of the SM theoretical calculation, the third column is the measured experimental value, the fourth is the 1$\sigma$ experimental error and the last column is this measured deviation between the SM prediction and the observed value from experiment.}
	\label{tab:QFV}
\end{table}
We show in Fig.~\ref{fig:my_label} the individual impact that each of the considered QFV observable poses to the points that have survived the electroweak and Higgs physics constraints.
The colour scheme is identical to that in \cref{fig:align,fig:MassP1} with the inclusion of green points that represent points simultaneously allowed by EW, Higgs and QFV observables. It is promptly noticeable that both $\epsilon_K$, panel c), and $\mathrm{BR}(B \rightarrow \chi_s\gamma)$, panel a), represent the weaker flavour physics constraints on the parameter space in such a way that all of the points that have survived the EW and Higgs physics constraints are also compatible with these two QFV observables. On the other hand, $\Delta M_s$, panel b), $\Delta M_d$, panel e), and $\mathrm{BR}(B_s\rightarrow \mu \mu)$, panel d), place a noticeable veto on the allowed parameter space. $\Delta M_s$ is the main culprit, with the lowest number of allowed points that survive such a constraint. We also note that, both $\Delta M_d$ and $\Delta M_s$ limit the mass to be below $1~\mathrm{TeV}$ for the $A_2$ pseudoscalar. Indeed, the low mass regime is where the vast majority of the allowed points tend to accumulate. For a cleaner analysis we shown in the histograms of \cref{fig:my_label1111} the number of points that are allowed by $\Delta M_s$, $\Delta M_d$, $\mathrm{BR}(B_s\rightarrow \mu \mu)$ and combinations of them, in bins of the lightest pseudoscalar mass.
We also show in \cref{tab:AR} the acceptance ratio, defined as the percentage of the initial sample consistent with EW and Higgs physics that also survives a certain QFV cut, for each of the considered QFV observables as well as the three combinations in \cref{fig:my_label1111}.
\begin{table}[htb!]    
	\centering
	\captionsetup{justification=raggedright,singlelinecheck=false}
	\begin{tabular}{c|c }
Set of QFV observables & Acceptance ratio \\[0.1CM] \hline
$\mathrm{BR}\left(B\rightarrow \chi_{s}\gamma \right)$ &   100.0\% \\[0.05CM]
$\mathrm{BR}\left(B_s\rightarrow \mu\mu \right)$ &  35.0\%  \\[0.05CM]
$\Delta M_d$ (GeV) & 48.0\%   \\[0.05CM]
$\Delta M_s$ (GeV) &  26.0\% \\[0.05CM]
$\epsilon_K$ (GeV) &  100.0\% \\[0.05CM]
{\color{blue}$\mathrm{BR}\left(B_s\rightarrow \mu\mu \right)$ \& $\Delta M_s$} &  {\color{blue} 9.39\%}   \\[0.05CM]
$\mathrm{BR}\left(B_s\rightarrow \mu\mu \right)$ \& $\Delta M_d$ &  22.21\%  \\[0.05CM]
$\Delta M_s$ \& $\Delta M_d$ &  25.57\%
\end{tabular}
	\caption{Acceptance ratio for each QFV (first five rows) and for pairs of QFV variables (last three rows). The most restrictive set is marked in {\color{blue} blue}.}
	\label{tab:AR}
\end{table}
Our results reveal that the sharpest razor of the parameter space is the combination of $\mathrm{BR}(B_s\rightarrow \mu \mu)$ and $\Delta M_s$ with a very low acceptance ratio of only 9.39\%, while their individual contributions are $35\%$ and $26\%$, respectively. This combination of observables also prefers pseudoscalar masses to be in a range between $125~\mathrm{GeV}$ and $1300~\mathrm{GeV}$. However, such a limited range must not be regarded as a particular feature of the $\nu$BGL-I model but merely as a result of the numerical scan, which is only efficient if all scalar masses are sufficiently light\footnote{When using an invertion procedure that receives as input parameters only physical masses and mixing angles, large hierarchies between the Higgs boson mass and the remaining scalars often result in unacceptably large values for the quartic couplings, which can be as large as several tenths of thousands.}. Indeed, it is well-known that multi-Higgs models typically observe a decoupling limit when the masses of the new scalars tend to infinity. Therefore, the study of parametric regions with large scalar masses requires a different strategy to increase the acceptance ratio as, \textit{e.g.}~it was done in \cite{Das:2021oik} by forcing the exact alignment limit and providing only the Higgs boson mass as input.
\begin{figure}[htb!]
    \centering
    \vspace{0cm}
    \captionsetup{justification=raggedright,singlelinecheck=false}
	\includegraphics[width=\textwidth]{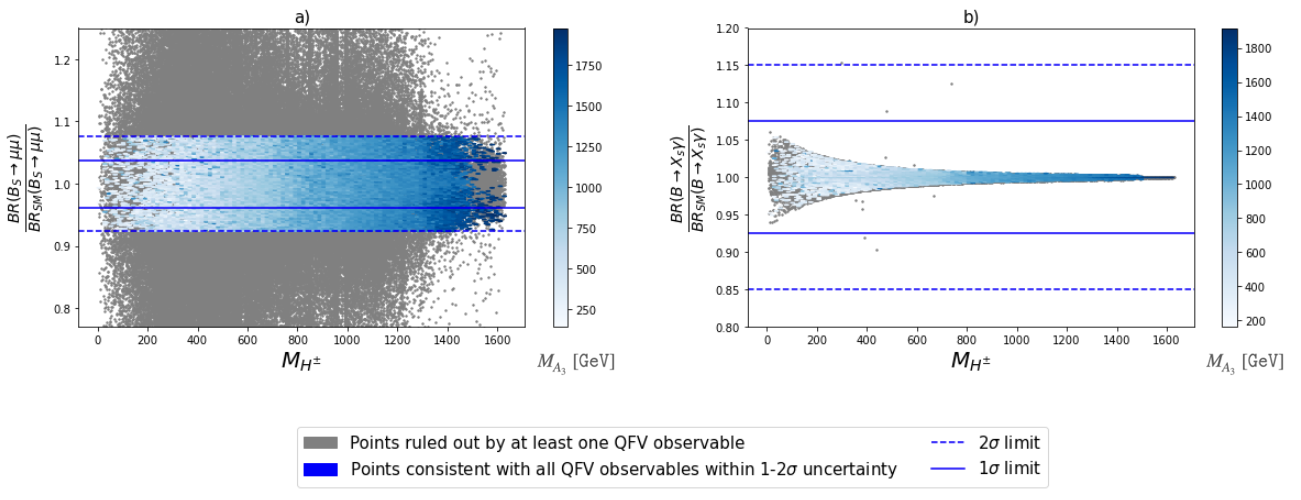}
	\caption{Plots of the ratio between the BR for the decays $B_S \rightarrow \mu^+\mu^-$ (on the left) and $B\rightarrow X_s \gamma$ (on the right) computed in the $\nu$BGL-I model and in the SM versus the mass of the charged scalar. In the colour bar, we showcase the mass of the heaviest pseudo-scalar $A_3$. Grey points are excluded from at least one QFV observable from \ref{tab:QFV} at a 2$\sigma$ uncertainty. Dashed blue lines indicate the $2\sigma$ bound and full blue lines indicate the 1$\sigma$ limit.}
	\label{fig:QFVcolor}
\end{figure}
\begin{figure}[htb!]
    \centering
    \vspace{0cm}
    \captionsetup{justification=raggedright,singlelinecheck=false}
	\scalebox{1}{\includegraphics[width=\textwidth]{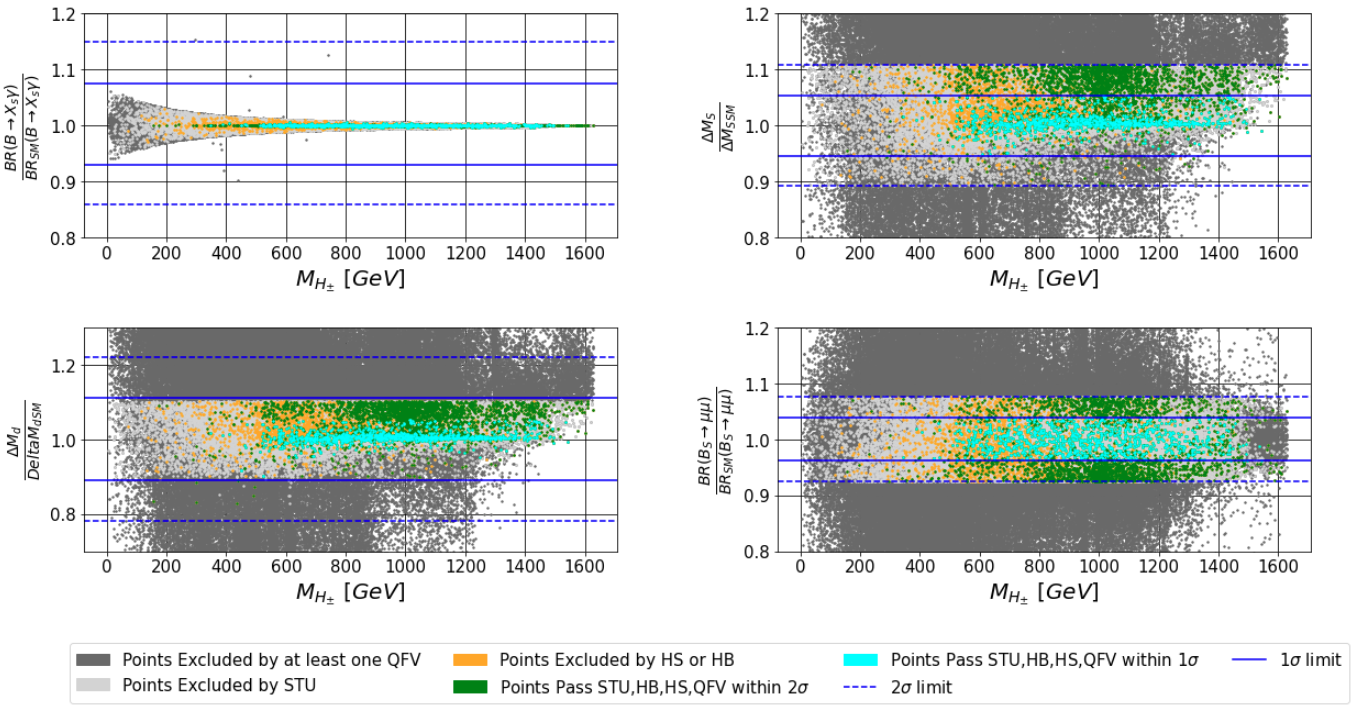}}
	\caption{Summary plots for the QFV variables. Dark grey points are excluded for at least one QFV observable in Table~\ref{tab:QFV} while light grey points do not pass the EW precision tests. Orange points are excluded either by HB or HS and green (cyan) points are compatible with EW and Higgs physics as well as all QFV constraints at a 2$\sigma$ (1$\sigma$) level.}
	\label{fig:QFVAll}
\end{figure}
For a closer inspection we show in \cref{fig:QFVcolor} two of the QFV observables of \cref{tab:QFV} in terms of the charged Higgs and heavier pseudoscalar masses. For a moment we ignore the EW and Higgs physics constraints and show in grey the points that are excluded by at least one QFV observable while the colored points are simultaneously allowed by all the QFV restrictions. Note that, even considering the entire universe of sampled points there are only two points outside the $\pm 2\sigma$ uncertainty band (dashed lines) that would fail $\mathrm{BR}\left(B\rightarrow \chi_{s}\gamma \right)$  (right panel) while $\mathrm{BR}\left(B_s\rightarrow \mu\mu \right)$ is visibly rather more restrictive (left panel). It is also clear that larger scalar masses tend to relax the QFV constraints as expected. 

Last but not least, we show in \cref{fig:QFVAll} the combination of all EW, Higgs and flavour physics constraints. We have considered the four mostly stringent QFV observables in \cref{tab:QFV}, leaving out only $\epsilon_K$ as it adds the least amount of constraints to our model. It is clear that the majority of sampled points fall in the dark grey area where at least one of the considered QFV observables lies beyond the $2\sigma$ uncertainty range limited by the blue dashed lines. As one can also see, flavour physics plays the major role in strongly reducing the allowed parameter space. In particular, for points complying with a $2\sigma$ uncertainty of all QFV observables (within the two blue dashed curves), we note that the majority of allowed cases also prefer the charged Higgs boson mass to be in a range between $400$ GeV and $1.4~\mathrm{TeV}$. If we further demand that all QFV observables are simultaneously reproduced within their 1$\sigma$ bound, then the viable region, identified by the cyan points, tends to be further compressed. Once again, it is important to mention that our scan was optimized to the low mass region, becoming increasingly inefficient with increasing scalar masses. Therefore one cannot extract bold conclusions about the high mass regime. On the contrary, what is important to retain as a key result of this article is that the $\nu$BGL-I model is compatible with the current EW, Higgs and QFV observables in the limit where new scalar masses are near the EW-scale. This opens up the possibility for testing the model in a foreseeable future, potentially already at the LHC run-III, as discussed below.

%%%%%%%%%%%%%%%%%%%%%%%%%%%%%%%%%%%%%%%%%%%%%%%%%%%%%%
\section{Phenomenology at the LHC}\label{sec:LHC}
%%%%%%%%%%%%%%%%%%%%%%%%%%%%%%%%%%%%%%%%%%%%%%%%%%%%%%

In this section we compare the points that have survived all constraints with current direct searches for neutral scalars at the LHC, focusing on $\tau\tau$ \cite{ATLAS:2020zms}, $ZZ$~\cite{ATLAS:2020tlo} and $WW$~\cite{ATLAS:2017uhp} final states. An interface between \texttt{SPheno} and \texttt{MadGraph} \cite{Alwall:2014hca} allows one to determine the neutral scalar production cross section times its decay branching ratio to a given final state.

\begin{figure}[htb!]
    \centering
    \vspace{0cm}
    \captionsetup{justification=raggedright,singlelinecheck=false}
	\scalebox{1.}{\includegraphics[width=\textwidth]{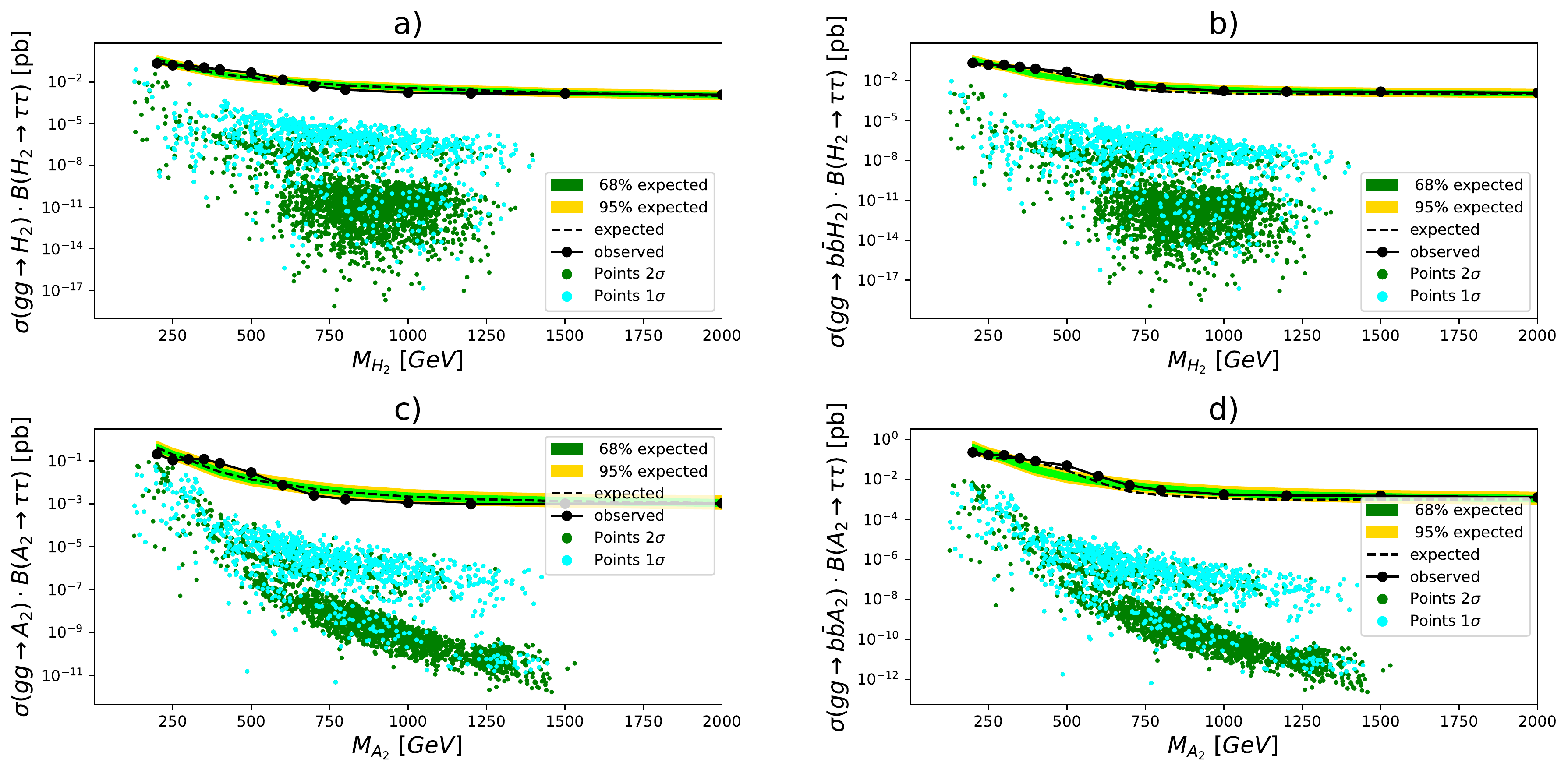}}
	\caption{The signal strength for production of a CP-even (top panels) and a CP-odd (bottom pannels) scalar via the gluon-gluon fusion mechanism (a), (c) times its branching ratio to $\tau \bar{\tau}$, and (b), (d) with associated production of a $b \bar{b}$-pair times its branching fraction to $\tau \bar{\tau}$, as a function of the lightest CP-even mass, $M_{H_2}$ (top panels) and lightest pseudoscalar mass $M_{A_2}$ (bottom panels). Green points satisfy QFV observables within 2 standard deviations, whereas blue ones pass QFV constraints at $1\sigma$. Green and yellow error bands correspond to the limits set by the ATLAS experiment \cite{ATLAS:2020zms} including run-II data collected at $\sqrt{s} = 13~\mathrm{TeV}$.}
	\label{fig:bra1}
\end{figure}

In \cref{fig:bra1} we show the latest ATLAS limits for scalar and pseudoscalar searches in the di-tau final state and confront them with our prediction for the lighter BSM Higgs bosons $H_2$ and $A_2$. The green (cyan) points have the same meaning as in \cref{fig:QFVAll}, \textit{i.e.}~ they are points for which the measured values of the considered QFV observables are reproduced in our calculation within a 2$\sigma$ (1$\sigma$) uncertainty. 
As one can observe, all points comfortably sit within the bounds set after the ATLAS run-II search. One may also take note that the calculated cross-sections in the $\nu$BGL-I model can become fairly small, even for masses close to the EW scale, although various points with sizable cross-sections are present. For the $H_2$ gluon-fusion production mechanism (\cref{fig:bra1}a), a point with the largest signal strength was found to be $0.083~\mathrm{pb}$, which can be probed in future LHC runs. An estimation of the expected number of events at the high-luminosity (HL) phase of the LHC, with $\mathcal{L}=3000~\mathrm{fb^{-1}}$, leads to $N=\sigma \mathcal{L}\sim 249078$, whereas for run-III luminosities one expects $N\sim 24907$. This suggests that the low $H_2$ mass regime can potentially be tested already with run-III data in the di-tau final state, as well as in the HL phase. For the case of gluon-fusion production in association with a $b \bar{b}$ pair (\cref{fig:bra1}b), we obtain slightly smaller values for $\sigma \cdot \mathrm{BR}$, where, the point with the highest cross-section is the same as in the previous process, with a mass of $132.27~\mathrm{GeV}$ and with the signal rate of $\sigma\cdot \mathrm{BR} = 0.0097~\mathrm{pb}$, which, at the (run III) HL phase, corresponds to an expected number of events of around $N\sim (2910)~29100$. As such, one can test this channel at both run-III and HL-LHC. A better picture is expected for the pseudoscalar production as one can see in the bottom panels of \cref{fig:bra1}, where one can find cross sections 1 to 2 orders of magnitude larger when compared to the analogous $H_2$ productions and decay channels.

The small signal strengths found for the di-tau final states are, to a large extent, a result of the small branching fractions generically observed for $H_2\rightarrow \tau^+\tau^-$ decays. Indeed, as one can see in \cref{tab:table-BRs}, for a selection of five benchmark points, the values of $\mathrm{BR}(H_2 \to \tau^+ \tau^-)$ are always sub-dominant, with the exception of benchmarks with sizable cross sections. 

Similar conclusions are obtained for the pseudoscalar case as one can see in the bottom plots of \cref{fig:bra1} and in the lower half of \cref{tab:table-BRs}. However, as mentioned above, $A_2$ production cross sections with subsequent decay into taus tend to be greater when compared to $H_2$, with the largest value encountered reaching $0.087~\mathrm{pb}$ for a pseudoscalar with mass $174~\mathrm{GeV}$. 

The benchmark points BP4 and BP2 in \cref{tab:table-BRs,tab:table-BRs-1} summarize the discussion of the previous two paragraphs, which were selected as the ones offering the larger $H_2$ and $A_2$ production and decay cross sections, respectively. For various displayed benchmarks, while for the $H_2$ field one finds multiple final states worth exploring, for the $A_2$ we note that quark decays are typically preferred. For the pseudoscalar case, we highlight benchmarks BP4 and BP5, which result in a final state with two tops and two charm quarks, respectively. In the later case, it would result in a final state with two light jets in the detector, whereas for the former case, one would expect at least two $b$-jets plus two $W$ bosons in the final state, following the decay of the top quarks.

Besides some points in the low-mass region, the small branching ratios are generically obtained for the scalar and pseudoscalar decays into a pair of taus, and as such one might wonder what then the dominant channels are. The answer is in part given in the last column of \cref{tab:table-BRs}, where we have found that $H_2$ mostly decays to $b\bar{b}$ and $WW$ pairs, as well as into $A_2\mathrm{Z^0}$ and two Higgs bosons, $H_1 H_1$. The first would result in a signature with two $b-$jets at the LHC while the second either in a pair of jets or a pair of leptons or even a jet and a lepton, together with missing energy. Note that the $H_2\rightarrow A_2\mathrm{Z^0}$ decay channel results in a interesting signature with at least four jets or two jets + two leptons, where we assume that the $A_2$ scalar mostly decays into c-jets, according to the discussion above. A more detailed analysis of this channel deserves special attention and its collider phenomenology is being the subject of a separate study currently in preparation. Notice that the $H_2 \to A_2 Z^0$ decay channel becomes increasingly important for heavier CP-even scalars as opposed to lighter $H_2$ scenarios where the pure SM decays dominate. One must also mention that $H_2$ can decay in a pair of Higgs bosons. Such a possibility is rather interesting as it can lead to a final state with $4b$-jets (via the chain $pp\rightarrow H_2 \rightarrow H_1 H_1 \rightarrow b\bar{b}b\bar{b}$). Indeed, both ATLAS and CMS have searched for Higgs boson pairs in the four b-jets final state~\cite{ATLAS:2018rnh,CMS:2018vjd,CMS:2017aza}. 

\begin{table}[htb!]    
	\centering
    \captionsetup{justification=raggedright,singlelinecheck=false}
	\resizebox{1.0\textwidth}{!}{\begin{tabular}{|c|c|c|c|c|c|c|}
		\toprule			
		\toprule
    	$\phi$ & ID & Mass (GeV) & $\mathrm{BR}(\phi \rightarrow \tau^+\tau^-)$ & $\sigma(gg\rightarrow \phi)\dotproduct \mathrm{BR}$ (pb) & $\sigma(gg\rightarrow b\bar{b}\phi)\dotproduct \mathrm{BR}$ (pb) & Maximum BR \\
		\hline
		\hline
		%\midrule
		\makecell{$H_2$}  & \makecell{BP1 \\ BP2 \\ BP3 \\ BP4 \\ BP5} &
		\makecell{$160.21$ \\ $347.99$ \\ $129.26$ \\ $132.27$ \\ $668.49$} &
		\makecell{$4.99\times 10^{-3}$ \\ $6.65\times 10^{-7}$ \\ $0.0127$ \\ $0.0357$ \\ $6.14\times 10^{-6}$} &
		\makecell{$0.007354$ \\ $3.25\times 10^{-8}$ \\ $3.59\times 10^{-4}$ \\  $0.0830$ \\ $7.42\times 10^{-8}$} &
		\makecell{$8.01\times 10^{-4}$ \\ $3.67\times 10^{-9}$ \\ $4.3\times 10^{-5}$ \\ $0.00967$ \\ $1.01\times 10^{-8}$} &
        \makecell{$\mathrm{BR}(W^+W^*) = 0.881$ \\ $\mathrm{BR}(H_1 H_1) = 0.611$ \\ $\mathrm{BR}(W^+ W^*) = 0.377$ \\ $\mathrm{BR}(b\bar{b}) = 0.590$ \\ $\mathrm{BR}(A_2 \mathrm{Z^0}) = 0.75$}
        \\
        \hline
        \hline
		\makecell{$A_2$}  & \makecell{BP1 \\ BP2 \\ BP3 \\ BP4\\ BP5} &
		\makecell{$194.99$ \\ $173.55$ \\ $1077.21$ \\ $937.61$ \\ $126.74$} &
		\makecell{$0.0336$ \\ $0.0264$\\ $2.65\times 10^{-5}$ \\ $3.49\times 10^{-5}$ \\ $2.48\times 10^{-3}$} &
		\makecell{$0.0546$ \\ $0.0874$ \\ $1.0\times 10^{-6}$ \\ $6.36\times 10^{-9}$ \\ $3.2\times 10^{-5}$} &
		\makecell{$0.00532$ \\ $0.008249$ \\ $1.62\times 10^{-7}$ \\ $8.66\times 10^{-10}$ \\ $3.0\times 10^{-6}$} &
		\makecell{$\mathrm{BR}(b\bar{b}) = 0.553$ \\ $\mathrm{BR}(b\bar{b}) = 0.432$ \\ $\mathrm{BR}(t\bar{t}) = 0.711$ \\ $\mathrm{BR}(t\bar{t}) = 0.918$ \\  $\mathrm{BR}(c\bar{c}) = 0.922$}
		\\
		\hline
	\end{tabular}}
	\caption{A selection of five benchmark points (BP) that respect all QFV, electroweak, Higgs and theoretical constraints. All masses are given in GeV and cross sections in pb. These correspond to $H_2$ early discovery scenarios that maximize the signal strength to di-bosons (BP1,BP2), and to later $H_2$ and $A_2$ discovery cases that maximize the di-tau channel (BP4) and again (BP2), respectively. The (BP3) represents the second lightest $M_{H_2}$ scenario while (BP5) was chosen to represent scenarios with heavier $H_2$ which preferably decay to $A_2 Z^0$. Di-tau final states are showcased while in the last column we provide the dominant decay mode with the respective branching ratio.}
	\label{tab:table-BRs}
\end{table}
\begin{table}[htb!]    
	\centering
    \captionsetup{justification=raggedright,singlelinecheck=false}
	\resizebox{1.0\textwidth}{!}{\begin{tabular}{|c|c|c|c|c|c|c|c|}
		\toprule			
		\toprule
    	$\phi$ & ID & Mass (GeV) & $\mathrm{BR}(\phi \rightarrow W^+W^-)$ & $\mathrm{BR}(\phi \rightarrow \mathrm{Z^0Z^0})$ & $\sigma(gg\rightarrow \phi \rightarrow W^+W^-)$ (pb) & $\sigma(gg\rightarrow \phi \rightarrow \mathrm{Z^0Z^0})$ (pb) & Maximum BR \\
		\hline
		\hline
		%\midrule
		\makecell{$H_2$}  &  \makecell{BP1 \\ BP2 \\ BP3 \\ BP4 \\ BP5} &
		\makecell{$160.21$ \\ $347.99$ \\ $129.26$ \\ $132.27$ \\ $668.49$} &
		\makecell{$0.881$  \\ $0.129$  \\ $0.377$   \\ $0.239$ \\ $0.416$} &
		\makecell{$0.0210$ \\ $0.0589$ \\ $0.0449$ \\ $0.0299$ \\ $0.203$} &
		\makecell{$1.0032$ \\ $0.00722$ \\ $3.0\times 10^{-6}$ \\ $0.00366$ \\ $0.00506$} &
		\makecell{$0.00054$ \\ $0.003319$ \\ $1.0\times 10^{-6}$ \\ $0.001426$ \\ $0.00244$} &
		\makecell{$\mathrm{BR}(W^+W^*) = 0.881$ \\ $\mathrm{BR}(H_1 H_1) = 0.611$ \\ $\mathrm{BR}(W^+ W^*) = 0.377$ \\ $\mathrm{BR}(b\bar{b}) = 0.590$ \\ $\mathrm{BR}(A_2 \mathrm{Z^0}) = 0.75$}
        \\
        \hline
        \hline
		\makecell{$H_3$}  & \makecell{BP1 \\ BP2 \\ BP3 \\ BP4 \\BP5} &
		\makecell{$929.20$ \\ $823.89$ \\ $1093.01$ \\ $1156.49$ \\ $754.64$} &
		\makecell{$0.124$ \\ $0.0922$ \\ $0.166$ \\ $0.0598$ \\ $0.0517$} &
		\makecell{$0.0616$ \\ $0.0455$ \\ $0.0824$ \\ $0.0297$ \\ $0.0254$} &
		\makecell{$0.00341$ \\ $1.36\times 10^{-4}$ \\ $0.002964$ \\ $0.000192$\\ $8.4\times 10^{-4}$} &
		\makecell{$0.00167$ \\ $6.7\times 10^{-5}$ \\ $0.001468$ \\ $9.4\times 10^{-5}$\\ $4.1\times 10^{-4}$} &
		\makecell{$\mathrm{BR}(H_1 H_1) = 0.456$ \\ $\mathrm{BR}(A_2 A_2) = 0.228$ \\ $\mathrm{BR}(H_1 H_2) = 0.531$ \\ $\mathrm{BR}(t\bar{t}) = 0.617$\\ $\mathrm{BR}(t\bar{t}) = 0.591$}
		\\
		\hline
	\end{tabular}}
	\caption{A selection of five benchmark points (BP) that respect all QFV, electroweak, Higgs and theoretical constraints. All masses are given in GeV and cross sections in pb. The benchmark points are the same as in \cref{tab:table-BRs}. Di-boson final states are showcased while in the last column we provide the dominant decay mode with the respective branching ratio.}
	\label{tab:table-BRs-1}
\end{table}

\begin{figure}[htb!]
    \centering
    \vspace{0cm}
    \captionsetup{justification=raggedright,singlelinecheck=false}
	\scalebox{1}{\includegraphics[width=\textwidth]{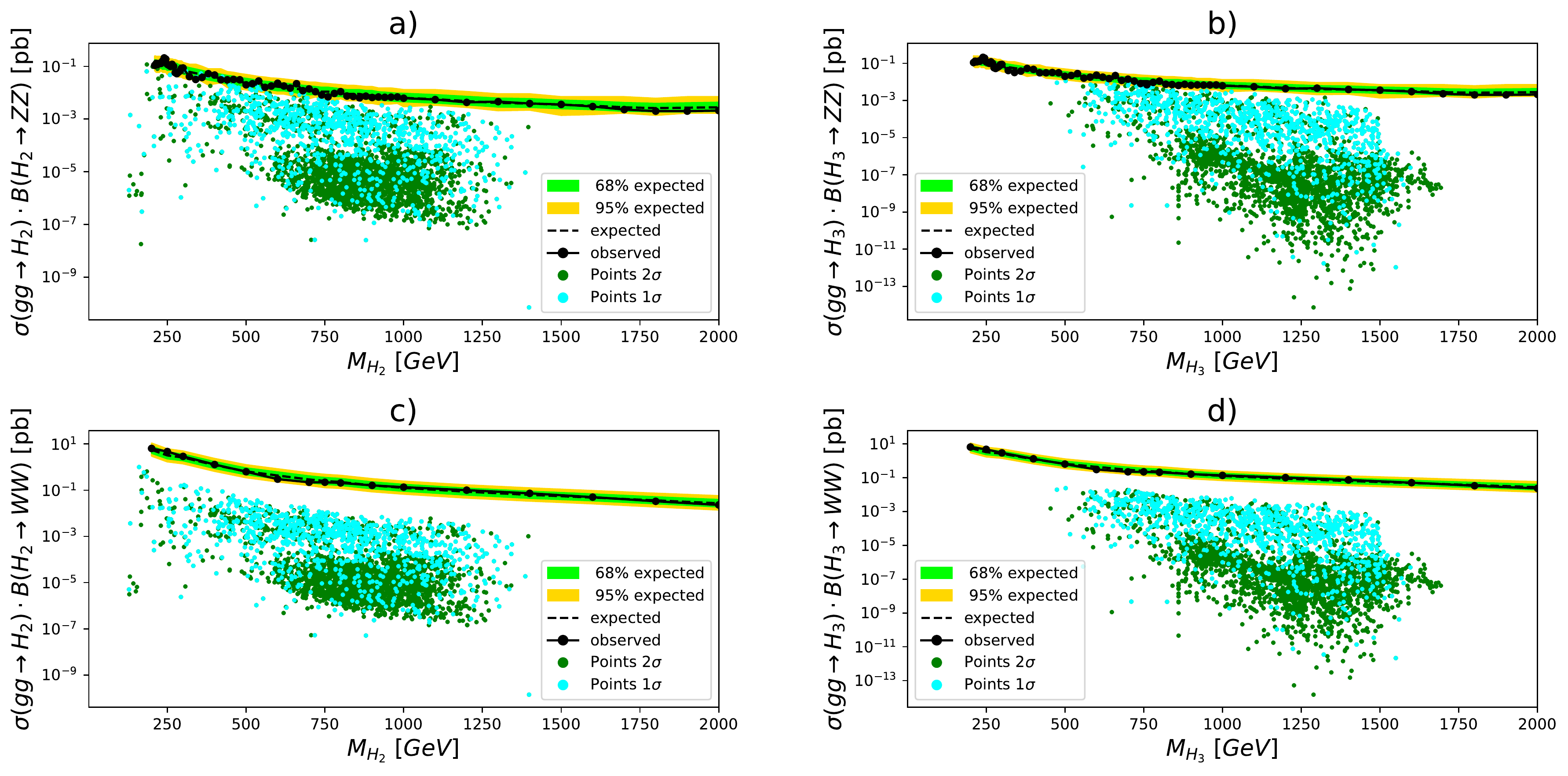}}
	\caption{The signal strength for production of a CP-even scalars via the gluon-gluon fusion mechanism (a), (b) times its branching ratio to $ZZ$, and (c), (d) times its branching fraction to $WW$, as a function of the lightest CP-even mass, $M_{H_2}$ (left panels) and heaviest CP-even mass $M_{H_3}$ (right panels). The colour code is the same as in \cref{fig:bra1} and the exclusion bounds were taken from \cite{ATLAS:2020tlo,ATLAS:2017uhp}.}
	\label{fig:brazz}
\end{figure}

In \cref{tab:table-BRs-1} one can also see that, even when not the dominant contributions, the decay of CP-even scalars into di-boson final states offer larger branching ratios than those of the tau channel. Recall that, in this article, we do not assume an exact Higgs alignment limit. Instead, we allow for small deviations from it, while still guaranteeing a SM-like Higgs candidate to have its mass and interactions within experimental bounds. In turn, the new CP-even scalars can couple to both $\mathrm{Z^0}$ and $W^\pm$ bosons and constraints arising from LHC direct searches, considering the $\phi\rightarrow W^+W^-$ and $\phi\rightarrow \mathrm{Z^0}\mathrm{Z^0}$ channels, with $\phi = H_2, H_3$, become relevant to our analysis. The results are summarized in the plots of \cref{fig:brazz}, for both the $H_2$ (left panels) and $H_3$ (right panels). As expected, all points remain bellow the experimental bounds but with larger signal strengths in comparison with the di-tau channel. In fact, most of points shown on panels a) and c) can be at the reach of the LHC run-III or its HL phase. In particular, we have found that for the diboson channel with $\mathrm{Z^0Z^0}$, a big portion of the allowed points comes close to the lower bound set by experiment, with the largest allowed cross-sections being of the order of $\mathcal{O}(10^{-1})$ and $\mathcal{O}(10^{-3})~\mathrm{pb}$. Therefore, it is possible the upcoming measurements would be capable of probing the allowed parameter space of the model. Note that for these two cases, the decay into a pair of W bosons is indeed the dominant channel. These points are very close to the ATLAS limit marking them as interesting benchmarks for an early probing, already with run-III data. It is also worth mentioning that, by virtue of a sizeable branching ratio to a W boson pair, the signal strength for the heavy CP-even state $H_3$ is comparable to those obtained for the lighter $H_2$ state. Finally, let us comment that, for the case of pseudoscalars, only the tau channel is considered given that tree-level couplings of $A_2$ to vector bosons are forbidden. Radiative effects where such processes can be induced are beyond the scope of our discussion in this article.

\section{Conclusions}
\label{sec:conclusions}

We have discussed anomaly-free implementations of a NTHDM-BGL model containing three generations of right-handed neutrinos, imbuing masses to the active left-handed neutrinos via a Type-I seesaw mechanism. We have specialized our numerical analysis on a version denoted as $\nu$BGL-I where flavour violation processes are expected only on the down-type quarks sector. We have constrained the allowed parameter space of the model upon imposing electroweak precision, Higgs and flavour physics constraints and scouted the parameter space in a search for phenomenologically valid regions. We have successfully assessed the viability of the low mass region and found that even for a number of scenarios with new scalars at, or at least very close to the EW scale, the $\nu$BGL-I model remains unconstrained. On the other hand, the majority of the excluded scenarios met their fate due to a combined effect of the $\Delta M_s$ and $\mathrm{BR}(B_s \rightarrow \mu \mu)$ QFV observables, which have eliminated approximately $90.61\%$ of the sampled points. 

For the points that have survived all constraints we have confronted our results with existing direct searches at the LHC. We have found that new CP-even scalars are largely favouring final states with a pair of W bosons for masses not far from the EW scale. However, as their mass grows, the decay channel $H_2 \to A_2 Z^0$ becomes dominant, thus a preferable option for further searches at the LHC run-III or HL phases. Last but not least, while our results confirm that the di-tau channel is well suited for pseudoscalar searches, their decay branching ratios to a pair of quarks is in general largely dominant.
\\ \\
%%%%%%%%%%%%%%%%%%%%%%%%%%%
{\bf Acknowledgements.}
%%%%%%%%%%%%%%%%%%%%%%%%%%%
%
Useful discussions and helpful correspondence with Mark Goodsell are gratefully acknowledged.
J.G., F.F.F., A.P.M. and V.V. are supported by the Center for Research and Development in Mathematics and Applications (CIDMA) through the Portuguese Foundation for Science and Technology (FCT - Funda\c{c}\~{a}o para a Ci\^{e}ncia e a Tecnologia), references UIDB/04106/2020 and UIDP/04106/2020. A.P.M., F.F.F., J.G. and V.V. are supported by the project PTDC/FIS-PAR/31000/2017. 
A.P.M.~is also supported by national funds (OE), through FCT, I.P., in the scope of the framework contract foreseen in the numbers 4, 5 and 6 of the article 23, of the Decree-Law 57/2016, of August 29, changed by Law 57/2017, of July 19.
J.G is also directly funded by FCT through a doctoral program grant with the reference 2021.04527.BD.
R.P.~is supported in part by the Swedish Research Council grant, contract number 2016-05996, as well as by the European Research Council (ERC) under the European Union's Horizon 2020 research and innovation programme (grant agreement No 668679). P.F. is
supported by CFTC-UL under FCT contracts UIDB/00618/2020, UIDP/00618/2020, and by
the projects CERN/FIS-PAR/0002/2017 and CERN/FIS-PAR/0014/2019.

\cleardoublepage

\appendix

\section{Anomaly conditions and charge assignment}\label{app:Conditions_charges}

The anomaly cancellation conditions independent of the neutrino flavour charges read as
\begin{align}\label{Anomaly_Cond}
\begin{split}
A_{_{\mathrm{U(1)_Y}\mathrm{U(1)_Y}\mathrm{U(1)^\prime}}}&\equiv\sum_{i=1}^{3}\left(
 X_{q_i} + 3X_{l_i} - 8X_{u_i} - 2X_{d_i} - 6X_{e_i} \right) = 0,\\
A_{_{\mathrm{U(1)_Y}\mathrm{U(1)^\prime}\mathrm{U(1)^\prime}}}&\equiv \sum_{i=1}^{3}\left(
 X_{q_i}^2 - X_{l_i}^2 - 2X_{u_i}^2 + X_{d_i}^2 + X_{e_i}^2\right) = 0,\\ 
A_{_{\mathrm{SU(2)_L}\mathrm{SU(2)_L}\mathrm{U(1)^\prime}}}&\equiv\sum_{i=1}^{3}\left(3X_{q_i} + X_{l_i} \right) = 0,\\
A_{_{\mathrm{SU(3)_C}\mathrm{SU(3)_C}\mathrm{U(1)^\prime}}}&\equiv\sum_{i=1}^{3}\left(2X_{q_i} - X_{u_i} - X_{d_i}\right) = 0\,,
\end{split}
\end{align}
with the $A_{XYZ}$ factor already defined in the main text below \cref{Anomaly_Cond_new}.

%\cleardoublepage
%===================================================

\section{Analytical equations for the Lagrangian parameters}\label{app:analytical_equations}
%===================================================
\begin{equation}\nonumber
\begin{aligned}
a_1 = & a_2+\frac{\left(M_{A_2}^2-M_{A_3}^2\right) \sin \left(2 \gamma _1\right)}{\sqrt{2} v},\\
\mu _3^2 = & \frac{M_{A_3}^2 \sin ^2\left(\gamma _1\right) \left(\cot \left(\gamma _1\right) v_S-v \sin (\beta ) \cos (\beta )\right)-M_{A_2}^2 \cos ^2\left(\gamma _1\right) \left(\tan \left(\gamma _1\right) v_S+v \sin (\beta ) \cos (\beta )\right)}{v}-\sqrt{2} a_2 v_S,\\
\mu _b^2 = &  -\frac{1}{2 v_S}\big(\sqrt{2} a_2 v^2 \sin (\beta ) \cos (\beta )+M_{A_3}^2 \cos \left(\gamma _1\right) \left(\cos \left(\gamma _1\right) v_S-v \sin (\beta ) \cos (\beta ) \sin \left(\gamma _1\right)\right)\\
&+M_{A_2}^2 \sin \left(\gamma _1\right) \left(\sin \left(\gamma _1\right) v_S+v \sin (\beta ) \cos (\beta ) \cos \left(\gamma _1\right)\right)\big)
, \end{aligned}
\end{equation}
%===================================================
\begin{equation}\nonumber
\begin{aligned}
\lambda _1= & \frac{1}{4 v^2}\big(\sec ^2(\beta ) \big(-\sin ^2(\beta ) \big(\big(M_{A_2}^2-M_{A_3}^2\big) \cos \big(2 \gamma _1\big)+M_{A_2}^2+M_{A_3}^2\big)\\
&+\sin \big(\alpha _2\big) \sin \big(2 \alpha _3\big) \big(M_{H_2}^2-M_{H_3}^2\big) \sin \big(2 \big(\alpha _1-\beta \big)\big)\\
&+2 \cos ^2\big(\beta -\alpha _1\big) \big(\cos ^2\big(\alpha _2\big) M_{H_1}^2+\sin ^2\big(\alpha _2\big) \big(\sin ^2\big(\alpha _3\big) M_{H_2}^2+\cos ^2\big(\alpha _3\big) M_{H_3}^2\big)\big)\\
&+2 \sin ^2\big(\beta -\alpha _1\big) \big(\sin ^2\big(\alpha _3\big) M_{H_3}^2+\cos ^2\big(\alpha _3\big) M_{H_2}^2\big)\big)\big),\\
\lambda _2 = & \frac{1}{4 v^2}\big(\csc ^2(\beta ) \big(-2 \cos ^2(\beta ) \big(M_{A_3}^2 \sin ^2\big(\gamma _1\big)+M_{A_2}^2 \cos ^2\big(\gamma _1\big)\big)\\
&+\sin \big(\alpha _2\big) \sin \big(2 \alpha _3\big) \big(M_{H_3}^2-M_{H_2}^2\big) \sin \big(2 \big(\alpha _1-\beta \big)\big)+2 \cos ^2\big(\alpha _2\big) M_{H_1}^2 \sin ^2\big(\beta -\alpha _1\big)\\
&+2 \sin ^2\big(\alpha _2\big) \sin ^2\big(\beta -\alpha _1\big) \big(\sin ^2\big(\alpha _3\big) M_{H_2}^2+\cos ^2\big(\alpha _3\big) M_{H_3}^2\big)\\
&+2 \cos ^2\big(\beta -\alpha _1\big) \big(\sin ^2\big(\alpha _3\big) M_{H_3}^2+\cos ^2\big(\alpha _3\big) M_{H_2}^2\big)\big)\big),\\
\lambda _3 =& \frac{1}{2 v^2}\big(\csc (\beta ) \sec (\beta ) \big(2 \sqrt{2} \big(a_1+a_2\big) v_S-\sin \big(2 \big(\alpha _1-\beta \big)\big) \big(\cos ^2\big(\alpha _2\big) M_{H_1}^2\\
&+\cos ^2\big(\alpha _3\big) \big(\sin ^2\big(\alpha _2\big) M_{H_3}^2-M_{H_2}^2\big)\big)+\sin \big(\alpha _2\big) \sin \big(2 \alpha _3\big) \big(M_{H_2}^2-M_{H_3}^2\big) \cos \big(2 \big(\alpha _1-\beta \big)\big)+4 \mu _3^2\big)\\
&+2 M_{A_3}^2 \sin ^2\big(\gamma _1\big)+2 M_{A_2}^2 \cos ^2\big(\gamma _1\big)+4 M_{H^{\text{pm}}}^2+\sin ^2\big(\alpha _3\big) \csc (2 \beta ) \sin \big(2 \big(\alpha _1-\beta \big)\big) \big(\cos \big(2 \alpha _2\big) M_{H_2}^2\\
&-M_{H_2}^2+2 M_{H_3}^2\big)\big),\\
\lambda _4 =&  -\frac{\csc (\beta ) \sec (\beta ) \left(\sqrt{2} \left(a_1+a_2\right) v_S+\sin (2 \beta ) M_{H^{\pm}}^2+2 \mu _3^2\right)}{v^2},
\end{aligned}
\end{equation}
%===================================================
\begin{equation}\nonumber
\begin{aligned}
\lambda _1^{\prime}= & \frac{1}{2 v_S^3}\big(v \sin (\beta ) \cos (\beta ) \big(\sqrt{2} a_2 v+\big(M_{A_2}^2-M_{A_3}^2\big) \sin \big(\gamma _1\big) \cos \big(\gamma _1\big)\big)+\sin ^2\big(\alpha _2\big) M_{H_1}^2 v_S\\
&+\cos ^2\big(\alpha _2\big) v_S \big(\sin ^2\big(\alpha _3\big) M_{H_2}^2+\cos ^2\big(\alpha _3\big) M_{H_3}^2\big)\big),\\
\lambda _2^{\prime} = & -\frac{1}{v v_S}\big(\sec (\beta ) \big(\sin (\beta ) \big(\sqrt{2} a_2 v+\big(M_{A_2}^2-M_{A_3}^2\big) \sin \big(\gamma _1\big) \cos \big(\gamma _1\big)\big)\\
&+\cos \big(\alpha _2\big) \big(\sin \big(\alpha _2\big) \cos \big(\beta -\alpha _1\big) \big(\sin ^2\big(\alpha _3\big) M_{H_2}^2+\cos ^2\big(\alpha _3\big) M_{H_3}^2-M_{H_1}^2\big)\\
&-\sin \big(\alpha _3\big) \cos \big(\alpha _3\big) \big(M_{H_2}^2-M_{H_3}^2\big) \sin \big(\beta -\alpha _1\big)\big)\big)\big),\\
\lambda _3^{\prime}=& -\frac{1}{v v_S}\big(\csc (\beta ) \big(\cos (\beta ) \big(\sqrt{2} a_2 v+\big(M_{A_2}^2-M_{A_3}^2\big) \sin \big(\gamma _1\big) \cos \big(\gamma _1\big)\big)\\
&+\cos \big(\alpha _2\big) \big(\sin \big(\alpha _3\big) \cos \big(\alpha _3\big) \big(M_{H_2}^2-M_{H_3}^2\big) \cos \big(\beta -\alpha _1\big)\\
&-\sin \big(\alpha _2\big) \sin \big(\beta -\alpha _1\big) \big(-\sin ^2\big(\alpha _3\big) M_{H_2}^2-\cos ^2\big(\alpha _3\big) M_{H_3}^2+M_{H_1}^2\big)\big)\big)\big)
\end{aligned}
\end{equation}
For completeness, in the case of a model with a neutral singlet (that is, with a $\mathrm{U(1)}^\prime$ charge of zero), all $a_1$, $a_2$, $a_3$ and $a_4$ would be simultaneously allowed, resulting in modified expressions for the parameters of the scalar potential apart from $\lambda^\prime_3$. These are
\begin{equation}\nonumber
\begin{aligned}
%----------------------------
a_4 =& \frac{1}{2 v v_S}\left( \left(-M_{A_2}^2 + M_{A_3}^2 \right)\sin{2\gamma_1} + \sqrt{2} v\left( a_1 - a_2 \right) + 2 v v_S a_3 \right)\,,\\
%----------------------------
\mu_3^2 =& \frac{1}{4 v}\Big( -v \left(M_{A_2}^2 +M_{A_3}^2 + \left( M_{A_2}^2 - M_{A_3}^2 \right) \cos{2\gamma_1}\sin{2\beta} \right)  + \left( M_{A_2}^2 - M_{A_3}^2 \right)v_s \sin{2\gamma_1} \\
 & + v v_S \left( -\sqrt{2}(3 a_1 + a_2) - 4 v_S a_3\right) \Big)\,,\\
%----------------------------
\mu_b^2 =& \frac{1}{4v_S} \Big( - \left(M_{A_2}^2 +M_{A_3}^2 \right)v_s + \left(M_{A_2}^2 - M_{A_3}^2 \right)\left(v_S\cos{2\gamma_1} + v\sin{2\beta}\sin{2\gamma_1}\right) \\
& + v^2 \cos{\beta}\sin{\beta}(-3 \sqrt{2}a_1 + \sqrt{2}a_2 - 8 v_S a_3) \Big)\,,
\end{aligned}
\end{equation}
\begin{equation}\nonumber
\begin{aligned}
%----------------------------
 \lambda_1 =&  -\frac{1}{2 v^2}  \Big( \sec{\beta}^2  \left(-M_{H_1}^2 \cos^2{\alpha_2} \cos^2{\delta} + \frac{1}{2} \left( M_{A_2}^2 + M_{A_3}^2 + (M_{A_2}^2 -M_{A_3}^2) \cos{2 \gamma_1}\right)\right)\sin^2{\beta} \\
 &-  M_{H_2}^2 \left(\cos{\delta} \sin{\alpha_2} \sin{\alpha_3} +\cos{\alpha_3} \sin{\delta}\right)^2 - M_{H_3}^2\left(\cos{\alpha_3} \cos{\delta}\sin{\alpha_2} -  \sin{\alpha_3} \sin{\delta}\right)^2 \Big)\,, 
 \end{aligned}
\end{equation}
\begin{equation}\nonumber
\begin{aligned}
%----------------------------
 \lambda_2 =&   
 \frac{1}{2 v^2} \csc^2{\beta} \Big(M_{H_3}^2 \cos^2{\delta} \sin^2{\alpha_3} - \cos^2{\beta} \left(M_{A_2}^2 \cos^2{\gamma_1} + M_{A_3}^2 \sin^2{\gamma_1}\right) \\
  &+ \left(M_{H_1}^2 \cos^2{\alpha_2} +  M_{H_2}^2 \sin^2{\alpha_2} \sin^2{\alpha_3}\right) \sin^2{\delta} + \cos^2{\alpha} \left( M_{H_2}^2 \cos^2{\delta} +  M_{H_3}\sin^2{a2} \sin^2{\delta}\right) \\
  &+ \left(-M_{H_2} + M_{H_3}^2\right) \cos{\alpha_3} \sin{\alpha_2} \sin{\alpha_3} \sin{2\delta}\Big)\,,
\end{aligned}
\end{equation}
\begin{equation}\nonumber
\begin{aligned}
%----------------------------
 \lambda_3 =& \frac{1}{8 v^2} \Big(-4 \left( M_{A_2}^2+ M_{A_3}^2 - 4 M_{H^{\pm}}^2 \right) 
 + 4 \left( - M_{A_2}^2 + M_{A_3}^2\right) \cos{2\gamma_1} \\
 & + \csc{\beta}\sec{\beta} \big(4 \left( M_{H_2}^2 - M_{H_3}^2 \right) \cos{2\delta} \sin{\alpha_2} \sin{2\alpha_3} + (2 \left(-2 M_{H_1}^2 + M_{H_2}^2 + M_{H_3}^2\right)\cos^2{\alpha_2} \\
 & - \left( M_{H_2}^2 - M_{H_3}^2\right) \left(-3 + \cos{2\alpha_2}\right) \cos{2\alpha_3}) \sin{2\delta}\big)\Big)\,,\\[0.5cm]
%----------------------------
 \lambda_4 =&\frac{1}{v^2} \left( M_{A_2}^2 + M_{A_3}^2 - 2 M_{H^{\pm}}^2 + \left( M_{A_2}^2 - M_{A_3}^2\right) \cos{2 \gamma_1}\right)\,,
 \end{aligned}
\end{equation}
\begin{equation}\nonumber
\begin{aligned}
 %----------------------------
 \lambda_{1}^\prime =&  \frac{1}{4 v_S^3}\left( 2 v_s \left(M_{H_1}^2 \sin^2{\alpha_2} +  \cos^2{\alpha_2}\left( M_{H_3}^2 \cos^2{\alpha_3} + M_{H_2}^2 \sin^2{\alpha_3}\right)\right) + 
  \sqrt{2} v^2 \cos{\beta} Sin{\beta} \left(a_1 + a_2\right)\right)\,,
\end{aligned}
\end{equation}
%%%%%%%%%%%%%%
\begin{equation}\nonumber
\begin{aligned}
%----------------------------
 \lambda_{2}^\prime =& \frac{1}{v v_S} \Big(\cos{\alpha_2} \sec{\beta} (\cos{\delta} \sin{\alpha_2} \left( M_{H_1}^2 - M_{H_3}^2 \cos^2{\alpha_3}
- M_{H_2}^2 \sin^2{\alpha_3}\right) \\
& + \left(-M_{H_2}^2 + M_{H_3}^2\right) \cos{\alpha_3}\sin{\alpha_3} \sin{\delta}) + \left(M_{A_2}^2 - 
       M_{A_3}^2\right) \cos{\gamma1} \sin{\gamma1} \tan{\beta} \\
       & - 
    v \left( \sqrt{2} a_1 + 2 v_S a_3\right) \tan{\beta}\Big)\,,
\end{aligned}
\end{equation}
%%%%%
and where the minimization conditions only modify $\mu_S^2$ and $\hat{\mu}$ as
$$
\mu_S^2 =\quad -\frac{\sqrt{2}v_1 v_2\left(a_1 +a_2\right)+v_S\left(v_1\left(2v_2\left(a_3 + a_4\right) + v_1\lambda_2^\prime\right) + v_2^2\lambda^\prime_3 +2\left(v_S^2\lambda^\prime_1 + \mu_{b}^2\right)\right) }{2v_S}
$$
and
$$
\hat{\mu} = v_S\left[\sqrt{2}\left(a_1+a_2\right) + v_S\left( a_3 + a_4\right)\right] + v_1 v_2\left(\lambda_3 + \lambda_4\right)+2\mu_3^2\,.
$$

\cleardoublepage

\bibliography{bib}

\end{document}